\documentclass[manuscript,screen,oneside,natbib=false]{acmart}

\usepackage[backend=biber, defernumbers=true, sortcites=true]{biblatex}
\usepackage{graphicx}
\usepackage{subcaption}
\usepackage{array}

\usepackage{booktabs}
\usepackage{multirow}
\usepackage{multicol}
\usepackage{makecell}
\usepackage{svg}
\usepackage[most]{tcolorbox}
\newtcolorbox{rqsummary}[1]{%
  colback=gray!10,
  colframe=black!80,
  boxrule=0.5pt,
  arc=2mm,
  left=1mm,
  right=1mm,
  top=1mm,
  bottom=1mm,
  fonttitle=\bfseries,
  title={#1}
}

\newtcolorbox{mygraybox}{
  colback=gray!20,
  colframe=gray!20,
  boxrule=0.5pt,
  arc=0pt,
  left=6pt, right=6pt, top=6pt, bottom=6pt
}

\RequirePackage[normalem]{ulem} 
\RequirePackage{color}
\definecolor{RED}{rgb}{1,0,0}
\definecolor{BLUE}{rgb}{0,0,1}

\providecommand{\DIFadd}[1]{\textcolor{blue}{#1}}
\providecommand{\DIFdel}[1]{\textcolor{red}{\sout{#1}}}
\providecommand{\DIFdel}[1]{\textcolor{blue}{\sout{#1}}}

\newif\ifshowdiffs
\showdiffsfalse 

\ifshowdiffs
  \renewcommand{\DIFadd}[1]{\textcolor{BLUE}{#1}}  
  \renewcommand{\DIFdel}[1]{}
\else
  \renewcommand{\DIFadd}[1]{#1}
  \renewcommand{\DIFdel}[1]{}
\fi

\usepackage{changepage} 

\usepackage{longtable}

\makeatletter
\def\@ACM@copyright@check@cc{}
\makeatother

\setcopyright{cc}
\setcctype{by}
\acmJournal{TOSEM}
\acmYear{2026} \acmVolume{1} \acmNumber{1} \acmArticle{}
\acmMonth{1} \acmDOI{10.1145/3809494}


\begin{document}

\title{The Impact of LLM-Assistants on Software Developer Productivity: \DIFadd{A Systematic Review and Mapping Study}}

\author{Amr Mohamed}
\orcid{0009-0000-6470-8970}
\email{amr.m@queensu.ca}
\affiliation{%
  \institution{Queen's University}
  \city{Kingston}
  \state{ON}
  \country{Canada}
}

\author{Maram Assi}
\email{assi.maram@uqam.ca}
\orcid{0000-0003-1274-7550}
\affiliation{%
  \institution{Université du Québec à Montréal}
  \city{Montréal}
  \state{QC}
  \country{Canada}
}

\author{Mariam Guizani}
\email{mariam.guizani@queensu.ca}
\orcid{0000-0003-2545-2612}
\affiliation{%
  \institution{Queen's University}
  \city{Kingston}
  \state{ON}
  \country{Canada}
}


\newcommand{\mariam}[1]{\textcolor{red}{{\it [Mariam: #1]}}}
\newcommand{\amr}[1]{\textcolor{blue}{{\it [Amr: #1]}}}
\newcommand{\Maram}[1]{\textcolor{purple}{{\it [Maram: #1]}}}
\begin{abstract}

Large language model assistants (LLM-assistants) present new opportunities to transform software development. Developers are increasingly adopting these tools across tasks, including coding, testing, debugging, documentation, and design. Yet, despite growing interest, there is no synthesis of how LLM-assistants affect software developer productivity. In this paper, we present \DIFadd{a systematic review and mapping of 39} peer-reviewed studies published between January 2014 and December 2024 that examine this impact. Our analysis reveals that \DIFadd{the majority of studies report considerable benefits from LLM-assistants, though a notable subset identifies critical risks}. Commonly reported gains include accelerated development, minimized code search, and the automation of trivial and repetitive tasks. However, studies also highlight concerns around cognitive offloading and reduced team collaboration. \DIFadd{Our study reveals that whether LLM-based assistants improve or degrade code quality remains unresolved, as existing studies report contradictory outcomes contingent on context and evaluation criteria}.  While the majority of studies (\DIFadd{90}\%) adopt a multi-dimensional perspective by examining at least two SPACE dimensions, reflecting increased awareness of the complexity of developer productivity, only \DIFadd{15}\% extend beyond three dimensions, indicating substantial room for more integrated evaluations. Satisfaction, Performance, and Efficiency are the most frequently investigated dimensions, whereas Communication and Activity remain underexplored. Most studies are exploratory (\DIFadd{59}\%) and methodologically diverse, but lack longitudinal and team-based evaluations. This review surfaces key research gaps and provides recommendations for future research and practice. All artifacts associated with this study are publicly available at \url{https://zenodo.org/records/18489222}.

\end{abstract}


\received{2 July 2025}
\received[revised]{4 Feb 2026}
\received[accepted]{17 March 2026}


\maketitle

\section{Introduction}
\label{sec:introduction}

Large Language Models (LLMs) are increasingly being integrated into the software engineering (SE) domain \cite{hou2024large}. In particular, the emergence of LLM-assistants, the term we use to refer to generative AI tools powered by LLMs that support software development tasks, has driven rapid adoption in both research and practice. Examples include OpenAI’s GPT-series (e.g., GPT-4 \cite{openai2023gpt4}) and GitHub Copilot \cite{githubcopilot}, which are now commonly used to assist with tasks such as code generation and completion\cite{buscemi2023comparative, gu2025effectiveness, kim2021code}, code translation \cite{zhang2025patch, weng2023automatic}, debugging and maintenance \cite{tian2024debugbench, dvivedi2024comparative, 10.1145/3744644}, documentation \cite{wang2024software, fan2023large}, and system design \cite{mandal2023large, white2024chatgpt}. These tools support a new development paradigm often referred to as AI pair programming, in which developers interactively engage with LLM-assistants throughout the software development process \cite{bird2022taking}. Since the public release of ChatGPT\footnote{\url{https://chatgpt.com/}} in late 2022, an expanding ecosystem of LLM-powered coding assistants—such as Cursor\footnote{\url{https://www.cursor.com/}}, Windsurf\footnote{\url{https://www.windsurf.com/}}, and Bolt\footnote{\url{https://bolt.new/}} has emerged. This widespread integration underscores the growing reliance on LLMs in SE and raises critical questions about their impact on software developer productivity.

Software developer productivity is a multifaceted construct that encompasses not only the efficiency and quality of software production but also the satisfaction, collaboration, and cognitive load experienced by a developer. While early approaches for measuring productivity relied on quantifiable outputs such as lines of code (LOC) or development velocity \cite{murphy2019predicts, beller2020mind}, recent research highlights the importance of human-centered factors such as communication, satisfaction, and well-being \cite{forsgren2021space}. As LLM-assistants become increasingly integrated into development workflows, it is crucial to understand their impact on software developer productivity.

To address this gap, we conduct a \DIFadd{systematic review and mapping of 39} peer-reviewed studies published between 2014 and December 2024 that examine the impact of LLM-assistants on software developer productivity. Our review analyzes methodological strategies, evaluation practices, and the productivity dimensions these studies engage with. We synthesize reported benefits and risks, and apply established conceptual frameworks to map our findings and contextualize their broader implications. Based on this synthesis, we identify key research gaps and provide actionable recommendations for both researchers and practitioners.
This paper makes the following contributions:

\begin{itemize}
\item \DIFadd{We present the first systematic review and mapping of the literature focused on the impact of LLM-assistants on software developer productivity, synthesizing evidence from 39} peer-reviewed primary studies published between 2014 and December 2024.

\item We provide a structured characterization of the methodological strategies and evaluation practices used to assess developer productivity, and synthesize the reported effects of LLM-assistants, surfacing key benefits (e.g., reduced task initiation overhead, support for code-adjacent tasks) and risks (e.g., over-reliance, flow disruption).

\item We analyze our findings through the lens of the \textit{SPACE} framework and employ McLuhan’s Tetrad framework in our discussion to reflect on broader socio-technical implications.

\item We offer actionable recommendations for practitioners and researchers, and release a publicly available replication package \cite{anonymous2025replication} containing all study data, selection decisions, and exclusion rationales to support transparency and reproducibility.

\end{itemize}

The remainder of this paper is structured as follows. Section ~\ref{sec:background} provides background on software developer productivity. Section ~\ref{sec:methodology} outlines the methodology used to conduct our review, including search strategies, selection criteria, and data extraction procedures. Sections ~\ref{sec:RQ0} through ~\ref{sec:RQ3} focus on addressing each of the research questions separately. Section ~\ref{sec:Discussion} discusses implications, recommendations for practitioners, and directions for future research. Section ~\ref{sec:threats} examines the threats to validity. Finally, section~\ref{sec:conclusion} concludes the paper.

\section{Background}
\label{sec:background}

\label{sec:Overview}

Software developer productivity has received sustained research attention since the early decades of software engineering. Notably, Frederick J Brooks noted in his 1975 book \textit{The Mythical Man Month} \cite{brooks1995mythical} that there is \textit{``no silver bullet''} for improving productivity, underscoring that software development is a complex, human-centric activity not easily optimized by simply adding more resources. \DIFdel{Since then, a wide range of studies have attempted to define and measure developer productivity \cite{forsgren2021space, mockus2002two, wagner2018systematic, graziotin2018happens, petersen2011measuring}.}

\DIFdel{However, despite decades of investigations, developer productivity remains a complex concept, and a clear consensus on how to define or measure it accurately has yet to be established. The lack of a universally accepted definition or precise measurement consensus is largely due to the wide range of variables that influence productivity, including individual factors (e.g., developer fluency, motivation, experience), organizational practices (e.g., team collaboration, feedback), and contextual elements (e.g., task variety, remote work capabilities) \cite{murphy2019predicts}. Traditionally, software developer productivity has been defined as the ratio of output to input \cite{petersen2011measuring, razzaq2024systematic,meyer2014software,meyer2017work,meyer2017characterizing}. Outputs have been quantified using metrics such as lines of code added \cite{mockus2002two, savor2016continuous}, tasks completed \cite{zhou2010developer}, or functions implemented \cite{jones1994software}, while inputs have been measured in terms of time spent (e.g., hours, days, or months) \cite{zhou2010developer, beller2025s}. Although such metrics offer a quantifiable view of developer output, they often fail to capture the broader human, social, and organizational dimensions that characterize real-world software development.}

This foundational perspective was further cemented by Tom DeMarco and Timothy Lister’s seminal work, \textit{Peopleware: Productive Projects and Teams} \cite{demarco2013peopleware}, which established that the biggest obstacles to software productivity are not technical, but rather sociological and organizational. The authors argue that improving developer performance requires managing people, environment, and culture, not just tools or processes. Similarly, work by Grinter et al. \cite{grinter1999geography} showed that the way software is architected and how teams are structured are deeply interconnected, misalignment between the two creates communication bottlenecks and coordination overhead. As projects scale or become divided across modules and teams, informal communication that once happened naturally becomes harder to sustain, requiring deliberate mechanisms for information sharing and decision tracking, illustrating that productivity is as much a social and organizational phenomenon as it is technical.

Since then, a wide range of studies have attempted to define and measure developer productivity \cite{mockus2002two, wagner2018systematic, graziotin2018happens, petersen2011measuring}. Despite decades of research, developer productivity remains a complex construct with no universal definition or measurement consensus. This ambiguity stems from the diverse variables that influence productivity \cite{murphy2019predicts}. Historically, software developer productivity has often been quantified using simple output-to-input ratios such as lines of code (LOC) \cite{mockus2002two, savor2016continuous} or task completion time \cite{zhou2010developer}. While these measures provide quantifiable proxies, they fail to capture the broader human, social, and organizational dimensions.


Recent research has shifted toward multidimensional frameworks to assess software developer productivity more comprehensively. This shift reflects the evolution of industry thinking toward continuous delivery and organizational performance \cite{forsgren2018accelerate}, which links technical capabilities, team culture, and delivery performance. Noda et al. \cite{noda2023devex} propose a model based on Developer Experience (\textit{DevEx})\cite{6225984}, comprising three core dimensions: feedback loops, cognitive load, and flow state. Their work synthesizes insights from software engineering and human-computer interaction to operationalize how development environments and processes shape developers’ day-to-day experiences. The \textit{DevEx} model offers a practical framework for assessing and improving productivity from the developer's point of view. Forsgren et al. \cite{forsgren2021space} introduce the \textit{SPACE} framework, which characterizes software developer productivity across five dimensions: Satisfaction and well-being, Performance, Activity, Communication and collaboration, Efficiency and flow. Satisfaction and well-being refer to how fulfilled and healthy developers feel in relation to their work, tools, team, and organizational culture. 
Performance reflects the quality and effectiveness of the outcomes of a system or process, such as software reliability, absence of bugs, or customer satisfaction. Activity captures the count of observable work events or software artifacts such as commits, pull requests, code reviews, builds, or deployments. Communication and collaboration address how individuals and teams communicate, coordinate, and integrate their work through metrics like discoverability of documentation and expertise. Efficiency and flow focus on the uninterrupted progress of work at both individual and system levels. Consequently, the \textit{SPACE} framework has been increasingly used in empirical studies to offer a more nuanced lens on productivity, especially in collaborative settings such as AI-assisted development \cite{smite2022changes, cheng2022improves, ruvimova2022exploratory, moe2021improving, storey2022developers, guenes2024impostor, vsmite2023forced}.

The suggested multidimensional perspectives highlight that a single metric cannot meaningfully capture productivity and instead encourage the use of composite measures that reflect the complex and varied nature of software development work, including human-centric metrics.

\section{Systematic Literature Review Methodology}
\label{sec:methodology}

We aim to establish the current state of evidence on the effects of LLM-assistants on software developer productivity. To this end, we conduct a comprehensive analysis across three key dimensions: (1) the methodological strategies, procedures, and instruments employed in primary studies (2) the reported benefits and risks associated with the use of LLM-based assistance, and (3) the specific dimensions of developer productivity that have been investigated. Our overarching objective is to synthesize the fragmented body of existing knowledge, highlight methodological strategies and their instrumentation, and identify critical gaps to guide future research in this rapidly evolving area. 

We ground our methodology in the seminal guidelines by Kitchenham and Charters \cite{keele2007guidelines}, which are derived from evidence-based practices in medical research and have been adapted for use in SE. 

In particular, this review is guided by the following research questions:\\

\noindent\textbf{RQ0: What are the characteristics of peer-reviewed studies that investigate the impact of LLM-assistants on software developer productivity?} 
\begin{adjustwidth}{0.8cm}{}
In RQ0, we contextualize the emerging research landscape surrounding LLM-assisted software development. Specifically, we examine this landscape from three angles: (1) an overview of the temporal distribution of publications, (2) the publication venues and their disciplinary focus, and (3) the patterns of authorship across the research community.\\
 \end{adjustwidth}

\noindent\textbf{RQ1: What are the methodological strategies, procedures, and instruments used by peer-reviewed studies that investigate the impact of LLM-assistants on software developer productivity?}  
\begin{adjustwidth}{0.8cm}{}
    In RQ1, we examine how existing research is conducted to investigate the relationship between LLM-assistants and software developer productivity. Specifically, we investigate the empirical strategies adopted to study the impact of LLM-assistants, (2) the procedures and study designs used to carry out these investigations, and (3) the instruments and metrics employed to evaluate software developer productivity.\\
 \end{adjustwidth}
 
\noindent\textbf{RQ2: What is the impact of LLM-assistants on software developer productivity?}
\begin{adjustwidth}{0.8cm}{}{
    In RQ2, we explore the effects of LLM-assistants on software development practices. First, we summarize the overall findings from the identified primary studies. We supplement these findings by analyzing the reported benefits and risks of using LLM-assistants across diverse study settings. This question aims to provide a structured understanding of how these tools affect developers in practice and what trade-offs they introduce.}\\

 \end{adjustwidth}

\noindent\textbf{RQ3: Which dimensions of developer productivity are investigated and how do these dimensions map onto the SPACE framework?} 
\begin{adjustwidth}{0.8cm}{} 
  
    In RQ3, we investigate how productivity is defined and assessed in the context of LLM-assisted software development. To provide additional insights, we map the main focus of each study to the dimensions of the \textit{SPACE} framework, i.e., Satisfaction and well-being, Performance, Activity, Communication and collaboration, and Efficiency and flow. This analysis offers a clear understanding of how the concept of productivity is operationalized across the existing body of work and highlights underexplored dimensions to guide future research.
     \end{adjustwidth}


\subsection{Pre-review mapping}

We adhere to the guidelines proposed by Kitchenham and Charters \cite{keele2007guidelines} for piloting the research protocol through pre-review mapping. To plan the construction of our review, we conducted a pre-review planning study that involved defining the research questions, establishing the inclusion and exclusion criteria, identifying a set of control papers, and iteratively refining the search strings. Detailed steps of this pre-review mapping (including the identification of control papers and query refinement) are all provided in the supplemental appendix \cite{anonymous2025replication}

\subsubsection{Control papers identification}

Once our research questions defined, we set our list of inclusion and exclusion criteria (see section~\ref{sec:inclusionexclusion}). We then performed a pilot search to identify a small set of control papers against which potential search string could be validated. This pilot involved manually searching for relevant publications, screening titles and abstracts, and conducting one round of backward and forward snowballing. The process yielded 17 control papers that met our criteria. These control papers, whose selection details are reported in the supplemental appendix \cite{anonymous2025replication}, were subsequently used to validate the search string. 

\begin{table}[h]
    \centering
    \caption{Database search strings and results. Total n = \DIFadd{9,756}.}
    \label{tab:search_strings}
    \resizebox{\textwidth}{!}{%
    \begin{tabular}{|l|p{12cm}|r|}
        \hline
        \textbf{Database} & \textbf{Search String} & \textbf{Results (since 2014)} \\
        \hline
        \textbf{ACM} & 
        (Language Model* OR “LM” OR “LMs” OR "LLM" OR “LLMs” OR "Artificial Intelligence" OR "AI")  
        AND  
        ((title:(Software Engineer* OR Software Develop* OR Developer* OR Coder* OR Programmer*))  
        OR  
        (abstract: (Software Engineer* OR Software Develop* OR Developer* OR Coder* OR Programmer*)))  
        AND (Productivity) & 4,044 \\
        \hline

        \textbf{IEEE Xplore} & 
        (((Language Model OR "LM" OR "LMs" OR "LLM" OR "LLMs" OR “Artificial Intelligence” OR "AI")  
        AND  
        (("Document Title":Software Engineer OR "Document Title":Software Develop* OR "Document Title":Developer OR "Document Title":Coder OR "Document Title":Programmer) NEAR/5 (Productivity))  
        OR  
        (("Abstract":Software Engineer OR "Abstract":Software Develop* OR "Abstract":Developer OR "Abstract":Coder OR "Abstract":Programmer) NEAR/5 (Productivity)))) & 491 \\
        \hline

        \textbf{ScienceDirect} & 
        ((Language Model OR “LM” OR “LMs” OR "LLM" OR "LLMs" OR "Artificial Intelligence" OR "AI") AND (Productivity))  \newline        
        Advanced Search: Title, abstract, keywords: (Software Engineer OR Software Development OR Developer OR Coder OR Programmer) & 3,734 \\
        \hline

        \textbf{Web of Science} & 
        ALL=(Language Model OR "LM" OR "LMs" OR "LLM" OR "LLMs" OR "Artificial Intelligence" OR "AI")  
        AND  
        ((TI=(Software Engineer NEAR Productivity OR Software Develop* NEAR Productivity OR Developer NEAR Productivity OR Coder NEAR Productivity OR Programmer NEAR Productivity))  
        OR  
        (AB=(Software Engineer NEAR Productivity OR Software Develop* NEAR Productivity OR Developer NEAR Productivity OR Coder NEAR Productivity OR Programmer NEAR Productivity))) & 271 \\
        
        \hline
        \DIFadd{\textbf{Scopus}} &  \DIFadd{ALL( LANGUAGE Model* OR "LM" OR "LMs" OR "LLM" OR "LLMs" OR "Artificial Intelligence" OR AI ) AND TITLE-ABS ( ( Software Engineer* W/5 Productivity ) OR (Software Develop* W/5 Productivity ) OR ( Developer* W/5 Productivity ) OR ( Coder* W/5 Productivity ) OR ( Programmer* W/5 Productivity ) )}
        & \DIFadd{836} \\
        \hline
        
        \DIFadd{\textbf{Springer}} & \DIFadd{(Language Model* OR "LM" OR "LMs" OR "LLM" OR "LLMs" OR "Artificial Intelligence" OR "AI") AND "Productivity"
        Title : (Software Engineer* OR Software Develop* OR Developer* OR Coder* OR Programmer*) }
        & \DIFadd{380} \\
        \hline

    \end{tabular}
    }
    \label{tab:search_strings}

\end{table}

\textbf{Inclusion and Exclusion Criteria.}
\label{sec:inclusionexclusion}
We select papers for our study based on the following inclusion (IC) and exclusion (EC) criteria. The supplemental material \cite{anonymous2025replication} includes the exclusion decision for each study based on specific IC \& EC criteria. 

\textbf{Inclusion (IC)}:
\begin{itemize}
    \item (\textbf{+}) IC1: The paper investigates the effect of AI or LLMs on software developer productivity.
    \item (\textbf{+}) IC2: The paper is in English.
    \item (\textbf{+}) IC3: The paper has an accessible full text and was published in 2014 or later.
\end{itemize}

\textbf{Exclusion (EC)}:
\begin{itemize}   
    \item (\textbf{--}) EC1 (Out of Scope): The paper does not focus on SE or does not explore the impact of AI or LLMs on software developer productivity.
    \item (\textbf{--}) EC2 (Out of Focus): The paper mentions the impact of AI or LLM on software developer productivity without it being one of the topics of the study.
    \item (\textbf{--}) EC3 (Publication Type): The paper belongs to any of the following categories: secondary studies; work-in-progress, extended abstracts, posters, tool demos, editorials, or grey literature; studies published in books, theses, workshop, monographs, keynotes, panels, doctoral symposium, or any other venues without a formal peer-review process.
    \item (\textbf{--}) EC4 (Length): The paper is a short publication with fewer than four pages.
    \item (\textbf{--}) \DIFadd{EC5 (Accessibility): The full text of the paper is not accessible online (e.g., behind paywalls without institutional access, unavailable PDFs, or inaccessible publisher archives).}

\end{itemize}


\subsubsection{Query formulation and refinement}
\label{sec:query_formation}

Following the guidelines proposed by Kitchenham and Charters \cite{keele2007guidelines}, \DIFadd{we selected six major digital libraries widely used in software engineering research: ACM Digital Library\footnote{\url{https://dl.acm.org/}}, IEEE Xplore\footnote{\url{https://ieeexplore.ieee.org/}}, ScienceDirect\footnote{\url{https://www.sciencedirect.com/}}, Web of Science\footnote{\url{https://www.webofscience.com/}}, SpringerLink\footnote{\url{https://link.springer.com/}}, and Scopus\footnote{\url{https://www.scopus.com/}}}.

Constructing effective search queries is a challenging step in developing the review protocol, particularly due to the absence of standardized guidelines for selecting search terms \cite{keele2007guidelines}. This process relies on iterative refinement \cite{macfarlane2022search}, especially in contexts where terminologies such as those used to describe AI and LLM in SE vary significantly between studies. Given the growing volume of research in this area, we observe that overly broad queries return a large number of false positives, thereby reducing the precision of the search query. Conversely, narrow queries risk omitting relevant studies \cite{zhang2011identifying}. We iteratively developed and refined the search string by validating candidate queries against the control papers, adjusting keywords and query structure as needed to balance precision and recall. After five query iterations, all authors held a consensus meeting and agreed on the final search string (see Table \ref{tab:search_strings}). The final selected search query successfully retrieves all 17 control papers

    \DIFdel{The first and last authors held five meetings to iteratively develop and refine the search string. In each iteration, we test candidate queries by checking whether they successfully retrieve all 17 control studies identified during the pilot phase \cite{zhang2011identifying} (see Figure~\ref{fig:ReviewPlanning}). Based on these tests, we adjust both the keywords and the structure of the queries to improve their relevance and coverage. For example, we refine the search strings by replacing terms like ``software engineering'' and ``software engineers'' with a wildcard term ``software engineer*'', following the syntax and capabilities of each database. We also avoid using the keywords ``assessment'', ``performance'', and ``efficiency'' after observing a high volume of irrelevant studies that focus on technical-level benchmarks rather than productivity. After five query iterations, all authors held a consensus meeting and agreed on the final search string (see Table \ref{tab:search_strings}). The final selected search query successfully retrieves all 17 control papers.}

Table~\ref{tab:search_strings} presents the finalized search query used in our review. Each search query consists of three segments separated by the \texttt{AND} string. The first segment related to AI or LLMs limits the filter to the AI or LLM technology. The second segment refers to the actor (i.e., software developers), and the third segment captures the concept in question (i.e., productivity).  

In alignment with recent SLRs ~\cite{li2025systematic, hamalainen2024systematic, lo2021systematic}, we restrict our searches to the title, abstract, and keywords, as this improves precision. We also leverage proximity operators in the query, namely ``NEAR/5'' \DIFadd{or ``w/5'' in IEEE Xplore and Scopus respectively}, to improve the contextual relevance of matched terms \cite{razzaq2024systematic}. This operator retrieves documents where the specified terms appear within five words of each other. \DIFadd{This refinement is only applied in IEEE Xplore, Web of Science and Scopus, as the ACM Digital Library, Springer and ScienceDirect do not support it.}

\subsection{Primary Study Selection Process }
\begin{figure}
    \centering    
    \includegraphics[width=0.8\textwidth]{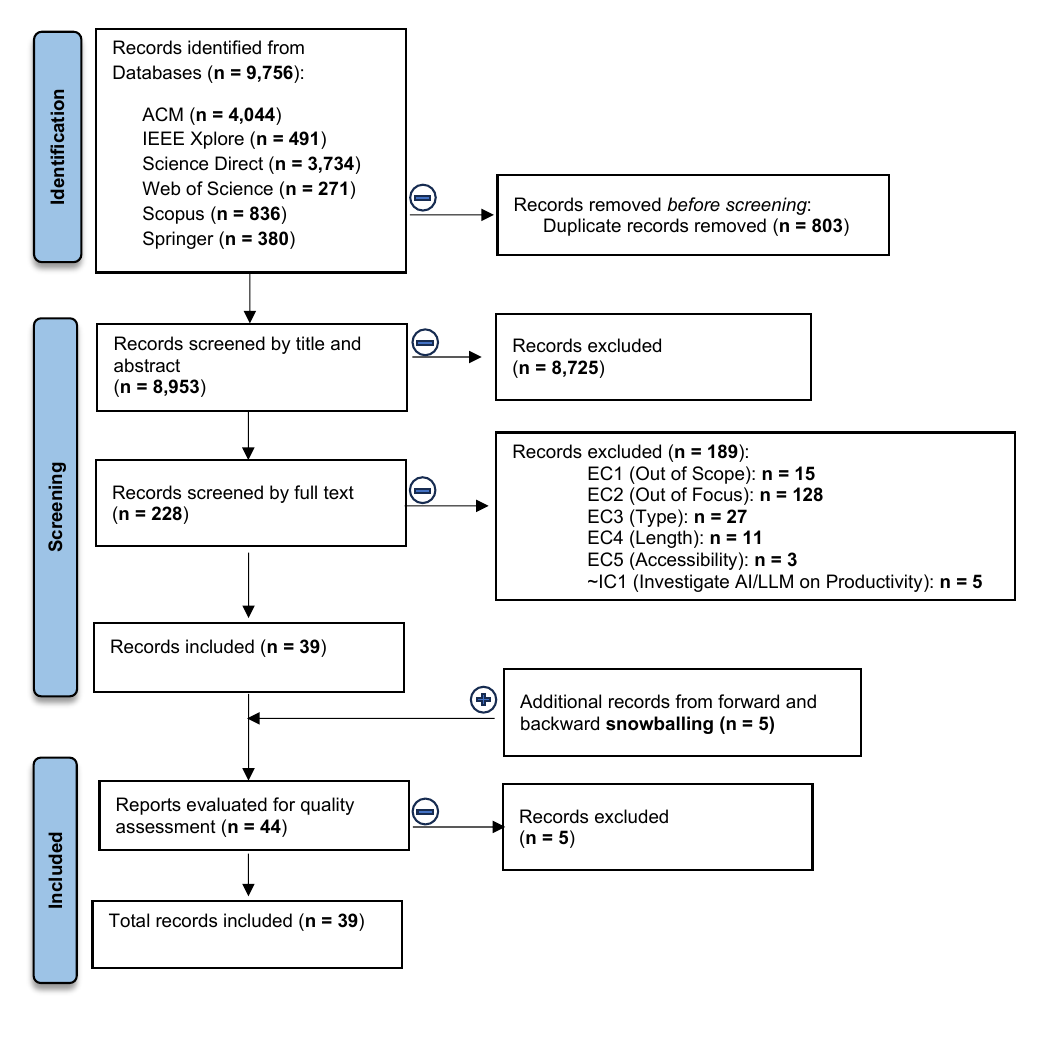}
    \caption{Overview of the selection process for primary studies included in the review using PRISMA flow chart \cite{PrismaFlowDiagram}.}
    \label{fig:prisma}
\end{figure}

We execute the search protocol as illustrated in Figure~\ref{fig:prisma}. First, we apply the finalized search strings on the selected digital libraries, restricting the results to publications from 2014 onward. This yields an initial set of \DIFadd{(n = 9,756)} records, including 4,044 papers from ACM, 491 papers from IEEE Xplore, 3,734 papers from ScienceDirect, 271 papers from Web of Science, \DIFadd{836 papers from Scopus, and 380 papers from Springer}. After removing duplicates, we obtain a set of \DIFadd{8,953} unique records (n = \DIFadd{803} excluded). The first author performs an initial screening of all records based on their titles and abstracts, a process that took a total of \DIFadd{47} days. We use the tool \textit{Rayyan}\footnote{\url{https://rayyan.ai/}} to tag any excluded paper with the corresponding exclusion criteria (the list of all excluded papers and their corresponding exclusion criteria is provided in the supplemental material \cite{anonymous2025replication}). The second and last authors independently validate the excluded papers. Three meetings were held to discuss any disagreements until reaching a consensus. We adopt a conservative screening approach whereby records with insufficient information in the title and abstract to support a clear inclusion or exclusion decision were included in full-text review. The title and abstract screening process resulted in \DIFadd{228} papers (see Figure~\ref{fig:prisma}) with a total of \DIFadd{8,725} excluded papers.

The full-text screening phase was conducted over a period of \DIFadd{10} weeks. This phase involves the careful reading and evaluation of the remaining \DIFadd{228} papers, applying the inclusion and exclusion criteria to assess their eligibility. For a detailed rationale on the inclusion and exclusion criteria during full-text screening, please refer to the supplemental material \cite{anonymous2025replication}. Throughout this process, the first author led the full-text screening of papers. Whenever the relevance of a study was unclear, it was flagged and reviewed in consultation with the second and last authors to ensure alignment with the protocol. This results in the exclusion of \DIFadd{189} studies during the full-text screening process.

To further expand our study set, we conduct a snowballing procedure by examining both the references and citations of the \DIFadd{39} selected studies. This phase took approximately two weeks, resulting in the identification and inclusion of five additional articles, expanding the set to \DIFadd{44} primary studies.
 
 \subsection{\DIFadd{Quality Assessment}}
\label{sec:quality_assessment}

\DIFadd{To ensure the reliability and rigor of the evidence synthesized in this review, we assessed the quality of the selected primary studies. We adopted the quality assessment strategy defined by Lenarduzzi et al.~\cite{lenarduzzi2021systematic}, which provides a comprehensive framework for evaluating empirical software engineering studies.}

\DIFadd{We evaluated each study against 11 criteria (QA1--QA11) designed to assess the clarity of research aims, the appropriateness of the methodology, the rigor of data analysis, and the validity of the findings. Table~\ref{tab:qa_criteria} details the specific criteria used.}

\begin{table}[h]
    \centering
    \caption{\DIFadd{Quality Assessment Criteria (adapted from Lenarduzzi et al. \cite{lenarduzzi2021systematic})}}
    \label{tab:qa_criteria}
    \begin{tabular}{|l|p{10cm}|}
        \hline
        \DIFadd{\textbf{ID}} & \DIFadd{\textbf{Quality Assessment Criterion}} \\
        \hline
        \DIFadd{QA1} & \DIFadd{Is the paper based on research (or is it merely a ``lessons learned'' report based on expert opinion)?} \\
        \hline
        \DIFadd{QA2} & \DIFadd{Is there a clear statement of the aims of the research?} \\
        \hline
        \DIFadd{QA3} & \DIFadd{Is there an adequate description of the context in which the research was carried out?} \\
        \hline
        \DIFadd{QA4} & \DIFadd{Was the research design appropriate to address the aims of the research?} \\
        \hline
        \DIFadd{QA5} & \DIFadd{Was the recruitment strategy appropriate for the aims of the research?} \\
        \hline
        \DIFadd{QA6} & \DIFadd{Was there a control group with which to compare treatments?} \\
        \hline
        \DIFadd{QA7} & \DIFadd{Was the data collected in a way that addressed the research issue?} \\
        \hline
        \DIFadd{QA8} & \DIFadd{Was the data analysis sufficiently rigorous?} \\
        \hline
        \DIFadd{QA9} & \DIFadd{Has the relationship between researcher and participants been considered to an adequate degree?} \\
        \hline
        \DIFadd{QA10} & \DIFadd{Is there a clear statement of findings?} \\
        \hline
        \DIFadd{QA11} & \DIFadd{Is the study of value for research or practice?} \\
        \hline
    \end{tabular}
\end{table}

\DIFadd{Following the scoring procedure proposed in \cite{lenarduzzi2021systematic}, each criterion was graded on a five-point Likert scale: \textit{Excellent} (4), \textit{Very Good} (3), \textit{Good} (2), \textit{Fair} (1), and \textit{Poor} (0).} \DIFadd{The detailed quality scores for each primary study are provided in the supplemental appendix~\cite{anonymous2025replication}.} \DIFadd{Studies that failed to meet a minimum quality threshold of 50\% average score \cite{garousi2016systematic} were excluded from the final set. This process resulted in the exclusion of 5 studies, leaving a final set of 39 primary studies. The majority of the studies were rated above 3. Detailed results of the quality assessment are available in the replication package material \cite{anonymous2025replication}.}

\subsection{Data Extraction and Synthesis}
In the final stage of the review process,  the synthesis of findings for RQs across these studies were conducted over a period of three months. we followed a qualitative synthesis approach, consistent with Kitchenham's guidelines \cite{keele2007guidelines}. To enable systematic integration, we first perform an initial thematic analysis iteration over all primary studies to extract relevant data (e.g., study goals, tools, empirical strategy and design, tasks, settings, and key results). In parallel, for each study, we write a descriptive summary. Then, our synthesis proceeded in multiple thematic analysis iterations. We first perform an initial thematic analysis phase to capture methodological details and observations. Then we follow up with three targeted iterations to capture methodological details (RQ1), synthesize benefits and risks (RQ2), and map studies to the SPACE framework (RQ3). After themes were consolidated, the first and last authors jointly validated the synthesized findings by cross-checking citations against the original text to ensure accuracy and traceability.

\section{RQ0: What are the characteristics of peer-reviewed studies that investigate the impact of LLM-assistants on software developer productivity?}
\label{sec:RQ0}

\subsection{Publication years }

\begin{figure}
    \centering
    \includegraphics[width=0.8\textwidth]{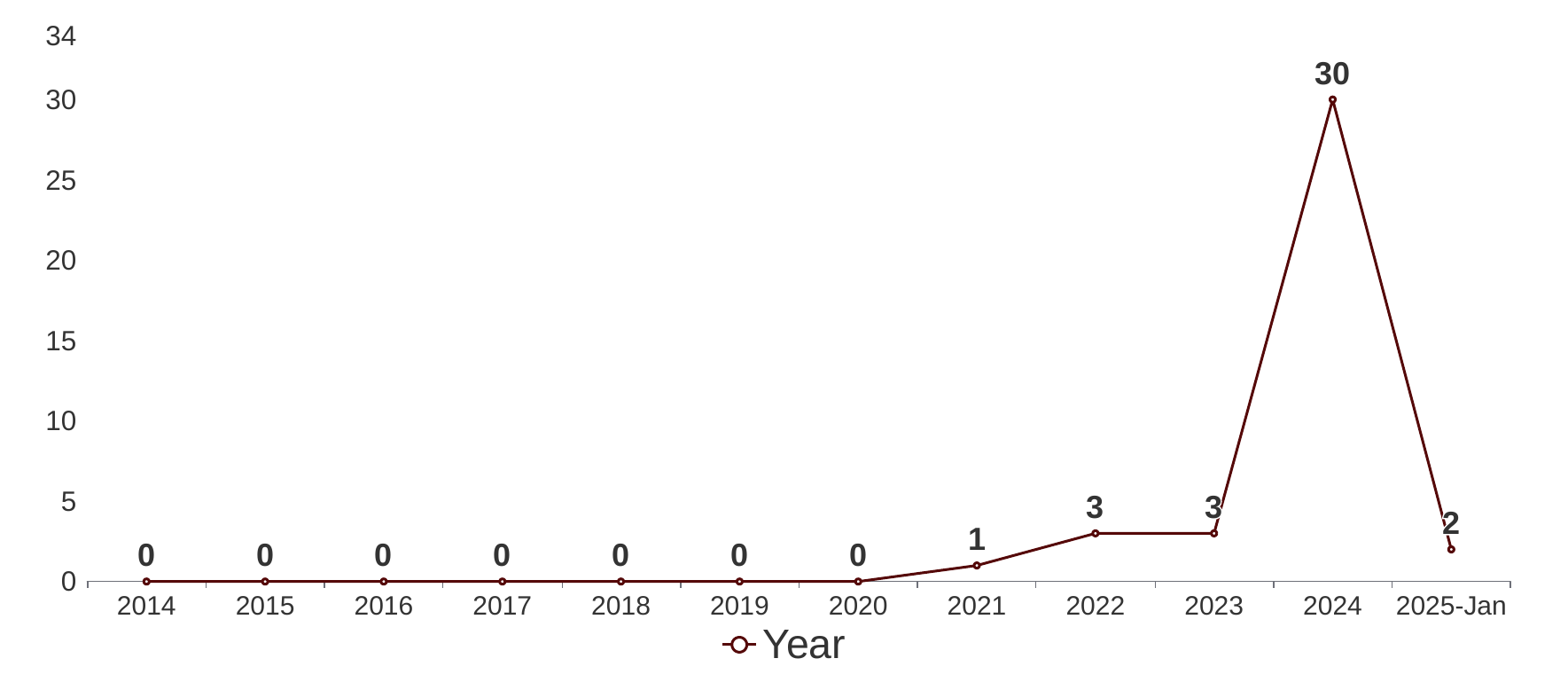}
    \caption{Publication frequency per year. }
    \label{fig:publicationsperyear}
\end{figure}

Figure~\ref{fig:publicationsperyear} shows the frequency of publications by year. Although we included studies published more than ten years preceding our review, only \DIFadd{four} were published between 2014 and 2022 \cite{PS23, PS10, PS37, PS65}. Research interest began to rise in 2022, which coincides with the release of ChatGPT\footnote{ChatGPT: \url{https://openai.com/chatgpt}}. This culminated in a sharp peak in 2024, which accounts for \DIFadd{77\%} of all included studies, likely due to increased accessibility and interest following ChatGPT’s release.

\subsection{Author distributions}
We investigate the distribution of papers per author across the \DIFadd{154} authors of primary studies. Most authors (i.e., \DIFadd{147}) have a single publication, while \DIFadd{6} authors have two papers, and the most prolific author, Igor Steinmacher, has three publications. This distribution is probably due to the fact that the investigation of the impact of LLM-assistants on software developer is a relatively new topic that is just starting to build momentum. 

\begin{table}[ht]
\centering
\caption{Distribution of primary studies by publication venues.}
\resizebox{\textwidth}{!}{%
\begin{tabular}{|p{4.2cm}|p{9.8cm}|p{2.8cm}|p{0.4cm}|}
\hline
\textbf{Research Focus} & \textbf{Venue} & \textbf{Primary Studies} & \textbf{\%} \\ \hline

\multirow{15}{*}{\parbox{4.2cm}{Software Engineering \newline and Computer Science}}
& Proceedings of the ACM on Software Engineering (PACMSE) & \cite{PS3, PS6, PS11, PS30} & \\ \cline{2-3}
& ACM Transactions on Software Engineering and Methodology (TOSEM) & \cite{PS23, PS25, PS48} & \multirow{15}{*}{} \\ \cline{2-3}
& International Conference on Software Engineering (ICSE) & \cite{PS2, PS22, PS40} & \\ \cline{2-3}
& Software Engineering in Practice (ICSE-SEIP) & \DIFadd{\cite{PS8}} & \\ \cline{2-3}
& ACM International Conference on the Foundations of Software Engineering (FSE) & \cite{PS69} & \\ \cline{2-3}
& ACM SIGPLAN International Symposium on Machine Programming (PLDI) & \cite{PS28} & \\ \cline{2-3}
& Automated Software Engineering (ASE) & \cite{PS9} & \DIFadd{46\%} \\ \cline{2-3}
& Software Quality, Reliability, and Security Companion (QRS-C) & \cite{PS51} & \\ \cline{2-3}
& Science of Computer Programming & \cite{PS43} & \\ \cline{2-3}
& Evaluation and Assessment in Software Engineering (EASE) & \cite{PS47} & \\ \cline{2-3}
& \DIFadd{International Conference on Evaluation of Novel Approaches to Software Engineering (ENASE)} & \DIFadd{\cite{PS61}} & \\ \hline

\multirow{5}{*}{\parbox{4.2cm}{Human-Computer \newline Interaction (HCI)}}
& ACM Conference on Human Factors in Computing Systems (CHI) & \cite{PS29, PS37} & \multirow{5}{*}{} \\ \cline{2-3}
& International Conference on Intelligent User Interfaces (IUI) & \cite{PS10, PS32} & \\ \cline{2-3}
& Proceedings of the ACM on Human-Computer \newline Interaction (CSCW) & \cite{PS31} & \\ \cline{2-3}
& ACM Transactions on Computer-Human Interaction (TOCHI) & \cite{PS20} & \DIFadd{18\%} \\ \cline{2-3}
& \DIFadd{Topics in Cognitive Science} & \DIFadd{\cite{PS65}} & \\ \hline

\multirow{5}{*}{\parbox{4.2cm}{Information Systems \\ and Decision Science}}
& Journal of Decision Systems (JDS) & \cite{PS46} & \multirow{5}{*}{} \\ \cline{2-3}
& Proceedings of the Americas Conference on Information Systems (AMCIS) & \cite{PS55} & \\ \cline{2-3}
& Innovations in Software Engineering Conference (ISEC) & \cite{PS5} & \DIFadd{13\%} \\ \cline{2-3}
& International Conference on Decision Aid Sciences and Applications (DASA) & \cite{PS57} & \\ \cline{2-3}
& \DIFadd{Hawaii International Conference on System Sciences (HICSS)} & \DIFadd{\cite{PS59}} & \\ \hline

\multirow{4}{*}{\parbox{4.2cm}{Human-Aspects and \newline Socio-Economic Impact}}
& Futures & \cite{PS21} & \multirow{4}{*}{} \\ \cline{2-3}
& Structural Change and Economic Dynamics & \cite{PS42} & \\ \cline{2-3}
& Cooperative and Human Aspects of Software Engineering (CHASE) & \cite{PS1} & \DIFadd{10\%} \\ \cline{2-3}
& \DIFadd{Annual Conference of the South African Institute of Computer Scientists and Information Technologists (SAICSIT)} & \DIFadd{\cite{PS60}} & \\ \hline

\multirow{4}{*}{\parbox{4.2cm}{AI for Software \\/ Engineering AI Engineering}}
& AI Engineering - Software Engineering for AI (IEEE/ACM) & \cite{PS13} & \multirow{4}{*}{} \\ \cline{2-3}
& AI-Powered Software (AIware) & \cite{PS36} & \\ \cline{2-3}
& International Conference on Generative Artificial Intelligence and Information Security (GAIIS) & \cite{PS33} & \DIFadd{8\%} \\ \hline

\multirow{2}{*}{\parbox{4.2cm}{Software Engineering Education}}
& Innovation and Technology in Computer Science Education (ITiCSE) & \cite{PS27} & \multirow{3}{*}{} \\ \cline{2-3}
& ICSE Software Engineering Education and Training (ICSE-SEET) & \cite{PS4} & \DIFadd{5\%} \\ \hline

\end{tabular}
}
\label{tab:venues}
\end{table}

\subsection{Publication venues}
We extract the publication venue for each primary study. 
Table~\ref{tab:venues} shows the venues categorized by research focus. The majority (\DIFadd{46\%}) of primary studies fall under``Software Engineering and Computer Science'' published in venues including  TOSEM, ICSE, and EASE. Human-Computer Interaction (HCI) is the second most prominent research focus with \DIFadd{18\%} \DIFadd{(7 out of 39)} primary studies published in venues such as CHI and \DIFadd{IUI}. The remainder of the primary studies are similarly distributed among specialized venues focusing on AI for Software Engineering / AI Engineering, Software Engineering Education, Information Systems and Decision Science, and Human-Aspects and Socio-Economic Impact. The variety of publication venues highlights the breadth and depth of the topic under study. The integration of LLM-assistants into the software development workflow introduces important considerations related to usability, automation, interaction design, and developer behavior.

\subsection{Most frequently used LLM tools} 

Table~\ref{tab:llm_tools} summarizes all LLM-assistants used across the primary studies. The most frequently evaluated tools are \DIFadd{ChatGPT (15 studies), Github Copilot (14 studies) and followed by Tabnine (3 studies), GPT-4 (3 studies), CodeWhisperer (3 studies) and GPT-3.5 (2 studies). All Other tools were used only once, such as Claude, and Codex.}

\begin{table}[ht]
\small
\caption{Summary of all LLM tools used in the primary studies and their associated study IDs.}
\label{tab:llm_tools}
\centering
\begin{tabular}{|l|c|p{8cm}|}
\hline
\textbf{Tool} & \textbf{Count} & \textbf{Primary studies} \\
\hline
ChatGPT & 15 & \DIFadd{\cite{PS11,PS30,PS4,PS40,PS47,PS43,PS5,PS1,PS2,PS8,PS36,PS57,PS60,PS61,PS69}} \\
\hline
GitHub Copilot & 14 & \DIFadd{\cite{PS28,PS29,PS48,PS27,PS55,PS31,PS1,PS8,PS2,PS36,PS57,PS59,PS60,PS69}} \\
\hline
Tabnine & 3 & \DIFadd{\cite{PS48,PS2,PS69}} \\
\hline
GPT-4 & 3 & \DIFadd{\cite{PS22,PS57,PS61}} \\
\hline
CodeWhisperer &3 & \cite{PS48,PS2,PS8} \\
\hline
GPT-3.5 & 2 & \DIFadd{\cite{PS31,PS61}} \\
\hline
\DIFadd{Claude} & \DIFadd{1} & \DIFadd{\cite{PS69}} \\
\hline
Codex & 1 & \cite{PS20} \\
\hline
\DIFadd{Gemini} & \DIFadd{1} & \DIFadd{\cite{PS69}} \\
\hline
GPT-3 & 1 & \cite{PS51} \\ 
\hline 
Ansible Lightspeed & 1 & \cite{PS9} \\ 
\hline 
Bard & 1 & \cite{PS32} \\ 
\hline 
CodeGen2 (7B) & 1 & \cite{PS48} \\ 
\hline 
GILT & 1 & \cite{PS40} \\ 
\hline 
Internal tool: CodeCompose & 1 & \cite{PS3} \\ 
\hline 
NL2Code PyCharm plugin & 1 & \cite{PS23} \\ 
\hline 
StackSpotAI & 1 & \cite{PS13} \\ 
\hline 
StarCoder (7B) & 1 & \cite{PS48} \\ 
\hline 
TransCoder & 1 & \cite{PS10} \\ 
\hline 
aiXcoder & 1 & \cite{PS48} \\ 
\hline 
OpenAI API & 1 & \cite{PS36} \\ 
\hline 
Midjourney & 1 & \cite{PS36} \\ 
\hline
\end{tabular}
\end{table}

\begin{rqsummary}{RQ0- Summary}
    The majority of studies (\DIFadd{35 out of 39}) were published after the release of ChatGPT in November 2022. Only \DIFadd{four} earlier studies (2014–2022) precede this period and rely on pre-LLM paradigms. Most authors (\DIFadd{147 out of 154}) contributed a single publication, while \DIFadd{seven} authors published two or more papers. Most studies were published in Software Engineering and Computer Science venues (\DIFadd{18} studies), followed by Human-Computer Interaction venues (\DIFadd{7} studies).
    \end{rqsummary}

\section{RQ1: What are the methodological strategies, procedures, and instruments used by peer-reviewed studies that investigate the impact of LLM-assistants on software developer productivity?}
\label{sec:RQ1}

\subsection{Distribution of the research strategies}
\begin{table}[h]
\centering
\caption{Distribution of research strategies across the primary studies.}
\label{tab:strategy_distribution}
\resizebox{\textwidth}{!}{%
\begin{tabular}{|p{4cm}|p{8cm}|p{2cm}|}
\hline
\textbf{Strategy} & \textbf{Primary Study} & \textbf{Percent} \\
\hline
Field Study 
& \cite{PS1, PS3, PS8, PS9, PS11, PS33, PS36, PS55, PS69} 
& \DIFadd{23\%} \\
\hline
Field Experiment 
& \cite{PS4, PS59} 
& \DIFadd{5\%} \\
\hline
Experimental Simulation 
& \cite{PS5, PS29, PS43, PS57, PS61} 
& \DIFadd{13\%} \\
\hline
Laboratory Experiment 
& \cite{PS6, PS10, PS13, PS20, PS22, PS23, PS27, PS30, PS31, PS32, PS37, PS40, PS47, PS48, PS51} 
& \DIFadd{38\%} \\
\hline
Sample Study 
& \cite{PS2, PS25, PS28, PS42, PS46, PS65} 
& \DIFadd{15\%} \\
\hline
Judgment Study 
& \cite{PS21, PS60} 
& \DIFadd{5\%} \\
\hline
\end{tabular}
}
\end{table}

We classify the \DIFadd{39} primary studies based on \textcite{stol2018abc} taxonomy for empirical software engineering strategies (see Table~\ref{tab:strategy_distribution}), which is a taxonomy built to distinguish studies' strategies according to their level of obtrusiveness (e.g., control) and generalizability (e.g., realism). This classification offers a structured understanding of how the research community has approached the investigation of AI tools. 
Laboratory experiments are the most common strategy, used by \DIFadd{38}\% of the primary studies \DIFadd{(15 out of 39)}. Laboratory experiments rely on controlled environments to isolate the effects of LLM-assistants on specific development tasks. Field studies are the second most common strategy, representing \DIFadd{23}\% \DIFadd{(9 out of 39)} of the primary studies. Field studies prioritize ecological validity by observing developer behavior in real-world settings without the need for researcher intervention. Sample studies, typically large-scale surveys, account for \DIFadd{15}\% \DIFadd{(6 out of 39)}, they aim to capture broad trends and perceptions across diverse developer populations.

Other strategies are less frequent. Field experiments \DIFadd{(5\%, 2 out 39)} resemble field studies in that they occur in real-world settings. However, unlike field studies, researchers actively manipulate specific variables such as changes to a tool or process to evaluate their effects within the natural context. Experimental simulations \DIFadd{(13\%, 5 out of 39)} combine elements from both laboratory and field studies by replicating real-world scenarios within controlled environments to study certain phenomena under realistic but simulated conditions. Judgment studies \DIFadd{(5\%, 2 out of 39)}  involve collecting expert opinions in a structured manner in a series of interviews and questionnaires.

\begin{figure}
    \centering
    \includegraphics[width=1.0\textwidth]{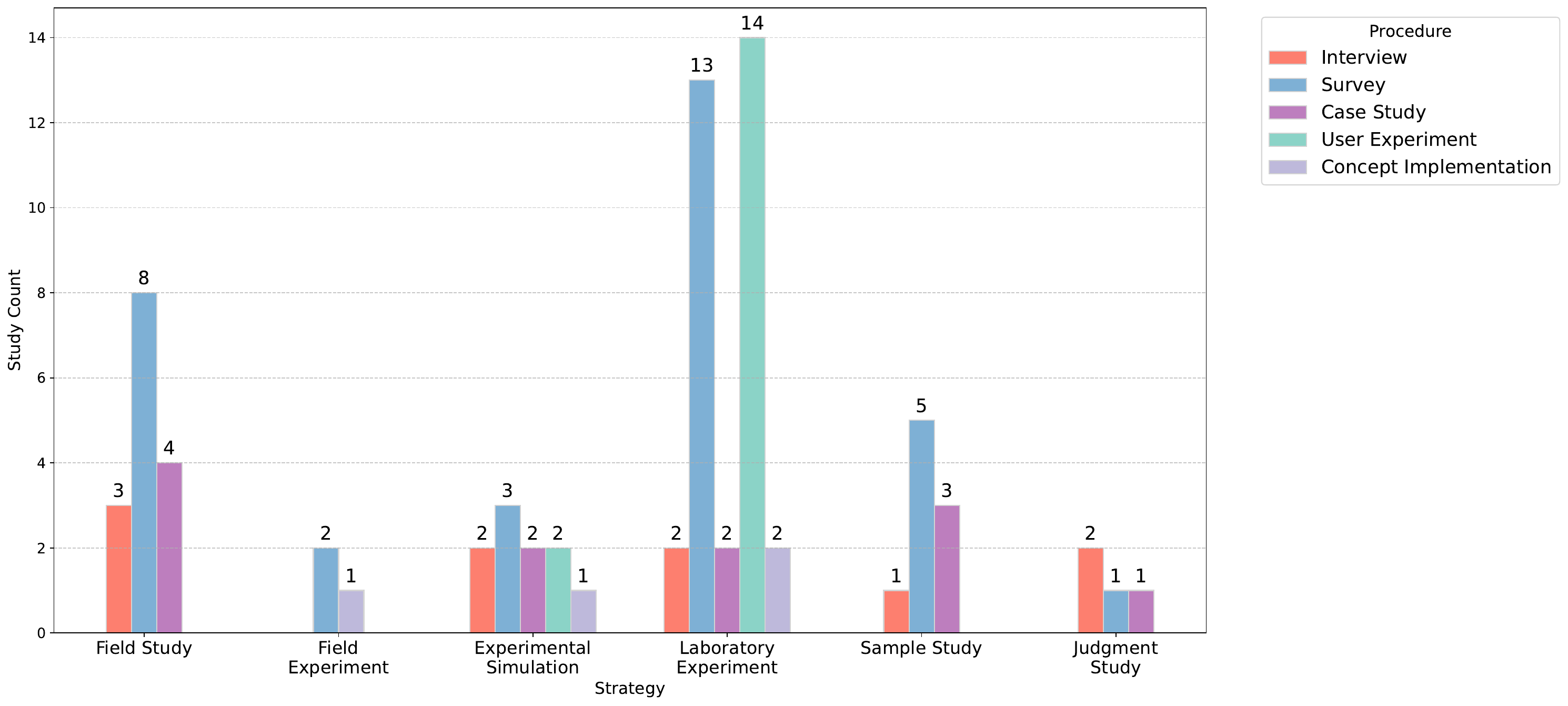}
    \caption{The distribution of empirical procedures across methodological strategies.}
    \label{fig:procedure_strategy_barchart}
\end{figure}

\subsection{Distribution of the research procedures}

\begin{table}[h]
    \centering
    \caption{Methodological procedures used in the empirical studies.}
    \resizebox{\textwidth}{!}{
    \begin{tabular}{|p{3.8cm}|p{11cm}|p{1.2cm}|}
        \hline
        \textbf{Procedure} & \textbf{Primary Studies} & \textbf{Percent} \\
        \hline
        Survey &
        \cite{PS2, PS3, PS4, PS8, PS9, PS10, PS11, PS13, PS20, PS21, PS22, PS23, PS25, PS27, PS28, PS29, PS30, PS31, PS32, PS33, PS36, PS37, PS40, PS42, PS43, PS46, PS47, PS51, PS55, PS59, PS61, PS69}
        & \DIFadd{82\%} \\
        \hline
        User Experiment &
        \cite{PS6, PS10, PS13, PS20, PS22, PS23, PS27, PS29, PS30, PS31, PS32, PS37, PS40, PS47, PS51, PS57}
        & \DIFadd{41\%} \\
        \hline
        Concept Implementation (Proof of Concept) &
        \cite{PS4, PS5, PS23, PS48}
        & \DIFadd{10\%} \\
        \hline
        Interview &
        \cite{PS1, PS20, PS21, PS33, PS37, PS46, PS57, PS60, PS61, PS69}
        & \DIFadd{26\%} \\
        \hline
        Case Study &
        \cite{PS5, PS8, PS10, PS28, PS33, PS36, PS37, PS46, PS55, PS60, PS61, PS65}
        & \DIFadd{31\%} \\
        \hline

    \end{tabular}
    }
    \label{tab:methodologies}
\end{table}


\begin{figure}
    \centering
    \includegraphics[width=0.9\textwidth]{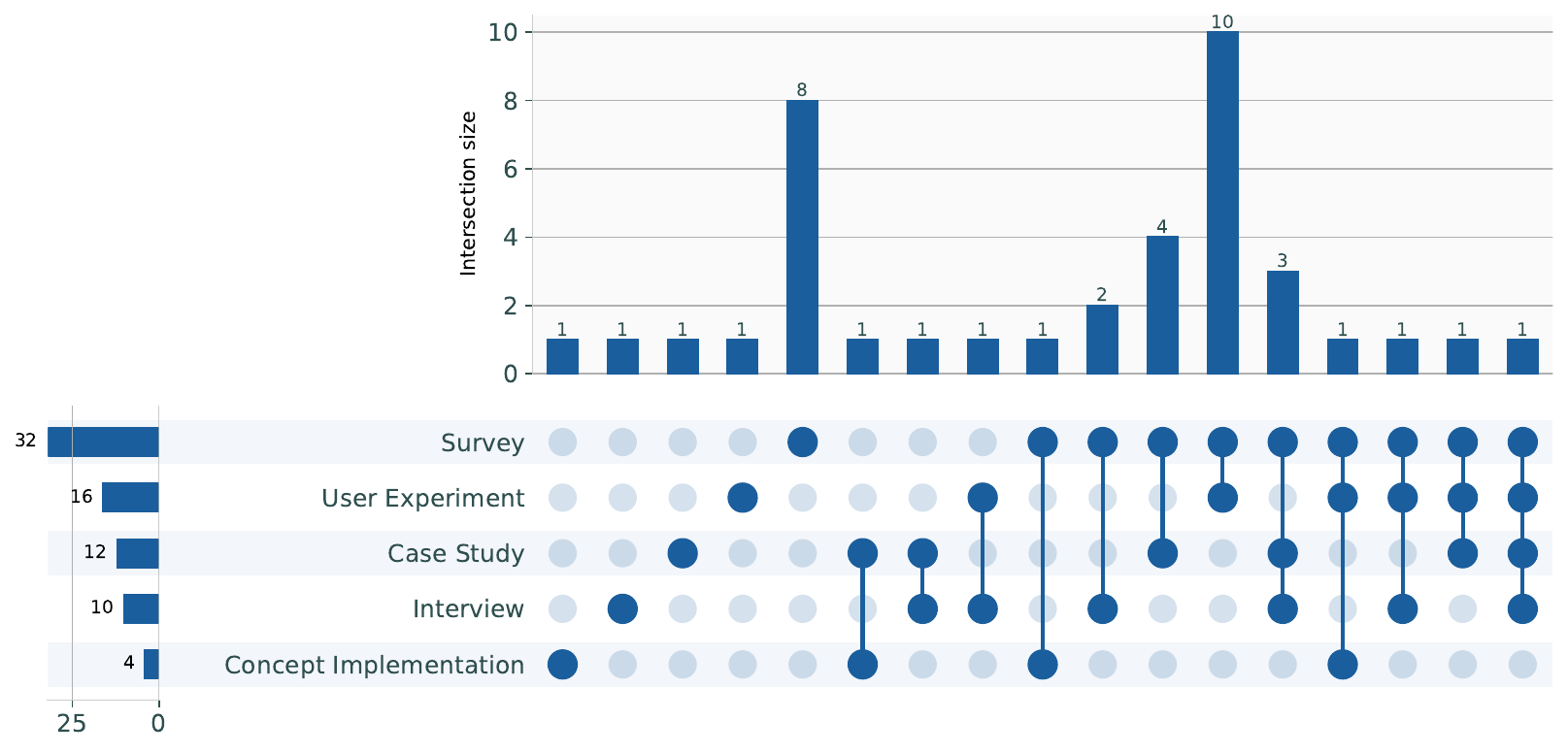}
    \caption{ Classification of methodological procedures classification and their overlap. Each bar on the left represents the number of studies that use a specific research method, while the bars along the top indicate how many studies employ a given combination of methods. For example, the most common pairing is a user experiment combined with a survey, used in 10 unique studies.}
    \label{fig:UpsetPlot}
\end{figure}

We adapt a fine-grained taxonomy identified from prior literature \textcite{glass2002research} to assign one or more methodology procedures to each primary study, such as interviews, surveys, or user experiments (i.e., controlled experiments, quasi-experiments), based on the primary methods used (see Table~\ref{tab:methodologies}). 

Among the primary studies, we find \DIFadd{90\% (35 out of 39)} of the studies leverage self-reported data such as surveys and interviews, and \DIFadd{41}\% of the studies (\DIFadd{16} out of \DIFadd{39}) are conducted in experimental settings (e.g., user studies, controlled experiments, or quasi-experiments). User experiments are almost exclusively associated with the laboratory experiment research strategy as shown in Figure~\ref{fig:procedure_strategy_barchart}. Additionally, \DIFadd{69}\% of these studies (\DIFadd{27} out of \DIFadd{39}) adopt a mixed-method approach as illustrated in Figure~\ref{fig:UpsetPlot}). For example, the most common combination of methods is a ``user experiment'' paired with a ``survey''. This approach is frequently used to triangulate self-reported perceptions (e.g., user experience or satisfaction) with measured performance metrics.

To provide additional insight into the nature of the current research direction, we assess the empirical studies based on their objectives, adopting a taxonomy from Hartson et al. \cite{hartson2001criteria} to classify the objective of each study into one of two approaches: formative and summative (see supplemental material \cite{anonymous2025replication} for detailed classification). A study with a formative objective primarily focuses on exploring, refining, or improving a process, tool, or methodology. In contrast, a study with a summative objective focuses on drawing conclusions about the effectiveness, outcomes, or impact of a completed process, tool, or methodology.  
We find that \DIFadd{59\% (23 out of 39)} of the studies have a formative goal, and \DIFadd{41\% (16 out of 39)} have a summative goal. This demonstrates that the current research direction reflect a balanced research landscape, rather than a strong focus on final, conclusive outcome.

We also classify each empirical study based on its adopted methods of data analysis: quantitative, qualitative, or both (see supplemental material \cite{anonymous2025replication} for classification details). \DIFadd{67\% (26 out of 39)} of the empirical primary studies include a mix of quantitative and qualitative analysis, \DIFadd{21\% (8 out of 39)} of the studies rely only on qualitative analysis, and \DIFadd{13\% (5 out of 39)} of the studies rely only on quantitative analysis.

\begin{table}[h]
\centering
\caption{Data sources and origin used by empirical studies.}
\resizebox{\textwidth}{!}{%
\begin{tabular}{|p{2.5cm}|p{3cm}|p{9cm}|} 
\hline
\textbf{Data Source} & \textbf{Instrument Origin} & \textbf{Instrument and Primary Studies} \\
\hline
\multirow{2}{*}{{\textbf{Self-Reported}}} 
& Designed by Authors 
& 
{Surveys} \DIFadd{\cite{PS2, PS4, PS8, PS10, PS11, PS21, PS23, PS25, PS30, PS31, PS32, PS33, PS36, PS43, PS47, PS51, PS57, PS61, PS69}} \newline
{Interviews} \DIFadd{\cite{PS1, PS20, PS21, PS33, PS37, PS46, PS57, PS60, PS61, PS69}} \newline
{Users open-ended feedback} \cite{PS3, PS9, PS13, PS20} \\
\cline{2-3}
& Validated Instruments and Frameworks 
& 
{NASA-TLX (Mental Effort)} \cite{PS10, PS20, PS22, PS27, PS30, PS40} \newline
{SPACE Framework-Based Surveys} \DIFadd{\cite{PS28, PS29, PS31, PS55}} \newline

{Technology Acceptance Model (TAM)} \cite{PS4, PS25, PS40} \newline
{Self-Efficacy Questionnaires} \cite{PS22, PS37} \newline
{After-Action Review for AI (AAR/AI)} \cite{PS22} \newline
{Emotion Affect Questionnaire} \cite{PS27} 
\\
\hline
\multirow{2}{*}{\parbox{2.5cm}{\centering \textbf{Behavioral\\\& Performance Metrics}}}
& Designed by Authors 
& 
{Task Completion and Correctness} \DIFadd{\cite{PS6, PS22, PS23, PS27, PS32, PS40, PS57, PS61, PS65}} \newline
{Suggestions Acceptance Rate} \cite{PS3, PS6, PS9, PS28, PS29, PS31, PS48} \newline
{Interaction Patterns (Logs/Edits/Tracking)} \cite{PS9, PS20, PS23, PS30, PS31, PS32, PS40} \newline
{Time to Completion} \DIFadd{\cite{PS5, PS6, PS20, PS23, PS30, PS31, PS32, PS37, PS40, PS51, PS55, PS65}} \newline
{Code Quality Metrics}\cite{PS10, PS22, PS23, PS31, PS33, PS51} \newline
{Productivity Gain} \DIFadd{\cite{PS5, PS57}} \\
\cline{2-3}
& Validated Frameworks 
& 
{Time Cost Quality (TCQ) Framework} \cite{PS42} \newline
{Resource-Based View (RBV) Framework} \cite{PS42, PS46} \newline
\\
\hline
\end{tabular}
}
\label{tab:instruments}
\end{table}

\subsection{Evaluation instruments}
\label{sec:InstrumentsUsed}
\begin{figure}[h]
        \centering
    \includegraphics[width=0.8\textwidth]{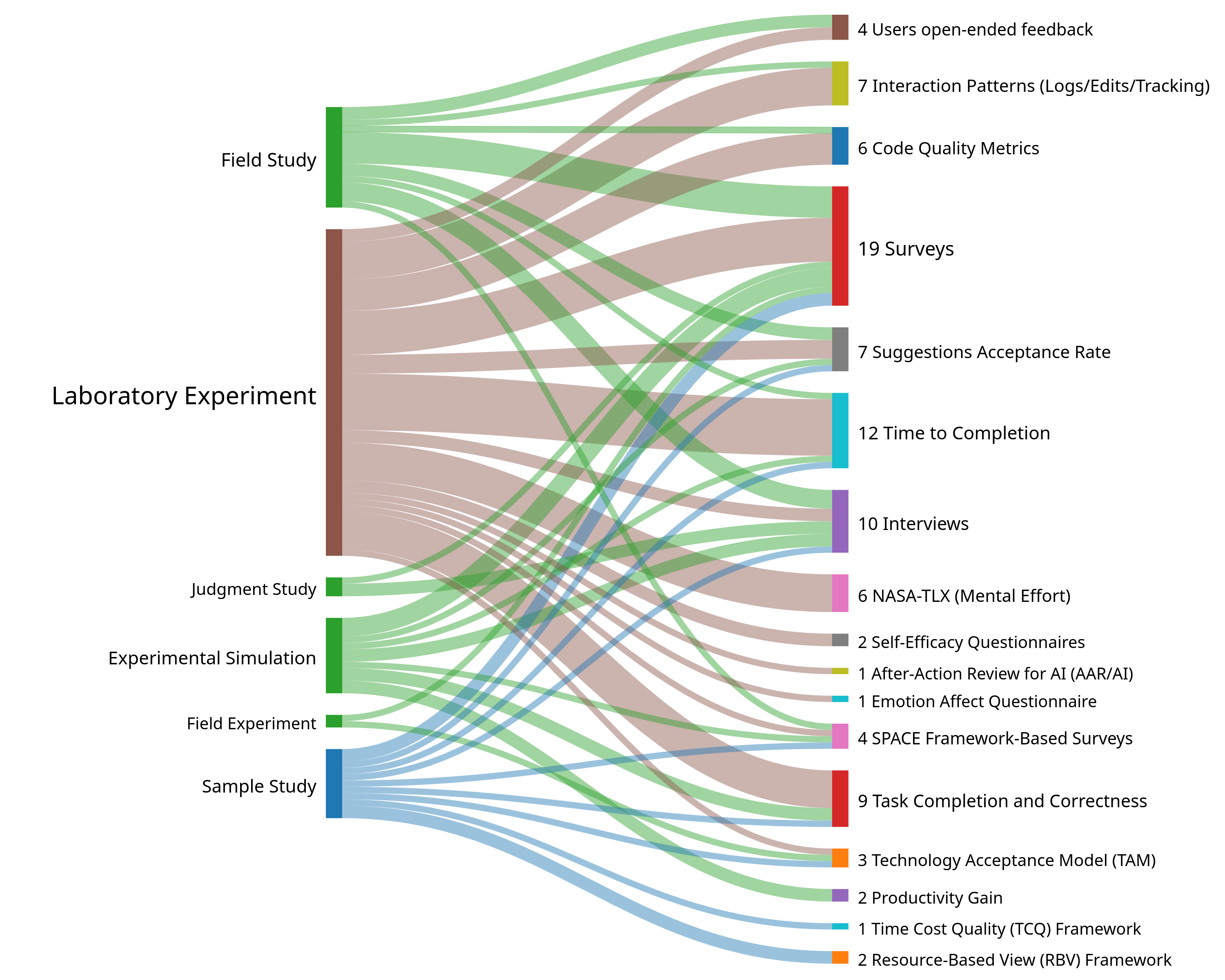}
        \caption{Mapping between empirical strategies and instruments in primary studies.}
    \label{fig:sankey_mappings}
\end{figure}

 Researchers employ various instruments, from self-reported surveys and interviews including validated questionnaires to behavioral \& performance metrics as shown in Table~\ref{tab:instruments}. 
 Self-reported methods remain predominant, often designed by study authors to capture user experience, perceived productivity, trust, or ease of use (e.g., post-task surveys, open-ended feedback). Only a subset of the studies \DIFadd{(15 out of 39)} incorporate validated instruments including the \textit{SPACE} framework \cite{forsgren2021space}, NASA-TLX for mental workload \cite{hart1988development}, TAM for technology acceptance \cite{silva2015davis}, self-efficacy questionnaires \cite{compeau1995computer, steinmacher2016overcoming}, or emotional affect questionnaire \cite{russell1980circumplex}. 
Behavioral \& performance metrics focus on quantifiable outcomes, such as time to completion, acceptance rate of AI-generated suggestions, code quality metrics, and some analysis of interaction patterns (see Table \ref{tab:instruments}). We find that behavioral \& performance metrics are mostly associated with studies with a high level of control, such as laboratory experiments, field experiments, or experimental simulation (see Figure~\ref{fig:sankey_mappings}), for example, metrics such as time to completion or code quality metrics are mainly associated with laboratory experiments. In contrast, field studies and sample studies, while still employing diverse sets of instruments, primarily rely on self-reported methods, such as surveys, interviews, and users' open-ended feedback.

\subsubsection{Time to completion}
\label{sec:timecompletion}
Measuring the time required to complete tasks is the most frequently used performance metric. \DIFadd{31\% (12 out of 39)} of the empirical primary studies employ this measure. The majority of studies that measured task completion time are laboratory experiments \DIFadd{\cite{PS6, PS20, PS23, PS30, PS31, PS32, PS37, PS40, PS51}}, which assess the time taken to complete specific programming tasks under controlled conditions. One field study conducted within a software company \cite{PS55} measures time-related performance by comparing throughput and cycle time before and after the integration of Copilot. One experimental simulation \cite{PS5} estimates total effort using a ``person/days'' metric to compare task completion durations with and without LLM-assistants. \DIFadd{Finally, one study \cite{PS65} measured completion time as the duration from when an issue was reported to when it was marked as resolved or closed.}

\subsubsection{LLM suggestions acceptance rate}
\label{sec:acceptancerate}
Measuring the acceptance rate of LLM suggestions is one of the commonly used behavioral and performance metrics to measure productivity \cite{PS3, PS6, PS9, PS28, PS29, PS31, PS48}. For instance, a study conducted by Meta \cite{PS3} evaluates the adoption of an internal LLM-assistant coding system (Code Compose) within the company. The study quantifies the LLM-assistant's utility by measuring both the number of LLM-generated suggestions accepted by developers and the proportion of code authored by the LLM-assistants. These metrics are then compared against reported acceptance rates from competing LLM-assistants to evaluate relative effectiveness.

One reason the acceptance rate metric is widely adopted is a study conducted by GitHub \cite{PS28}, which statistically analyzes the relationship between several interaction metrics related to code completion and developers’ self-reported productivity. The findings reveal a strong correlation between the frequency of accepted suggestions and perceived productivity. Despite these findings, the authors caution against using this metric in isolation to assess the effectiveness of LLM-assistants. They highlight that optimizing for acceptance rate may bias LLM-assistants toward well-represented languages or routine tasks, potentially disadvantaging less-represented workflows. Moreover, they warn that ``blind'' reliance on acceptance rate can lead to superficial improvements that inflate perceived usefulness without meaningfully enhancing developer outcomes.

\subsubsection{Mental effort and cognitive load}
\label{sec:cognitiveLoad}
Studies often use the terms mental effort or cognitive load interchangeably \cite{gonccales2021measuring}. Reducing mental effort is considered a motivation for incorporating LLM-assistants in software development. Traditionally, studies aim to measure developer cognitive load through biometric modalities, such as electroencephalogram (EEG) or electrocardiogram (ECG) \cite{gonccales2021measuring}. None of the identified studies leverages any biomedical measures or sensors for measuring cognitive load. Only one experimental study leverages eye-tracking \cite{PS6} to measure the time participants spend reading code documentation. 

Six studies (6 out of \DIFadd{39}) \cite{PS10, PS20, PS22, PS27, PS30, PS40} aim to measure cognitive workload by using NASA-TLX \cite{hart1988development}, which is a widely used questionnaire for assessing perceived mental workload. It captures six dimensions: mental demand, physical demand, temporal demand, performance, effort, and frustration. All six studies measure cognitive load in comparative experimental settings. We identify mixed findings regarding LLM-assistants' impact on mental cognitive load. In fact, a set of studies reports improvements \cite{PS27, PS32, PS40}, others neutral effects \cite{PS10, PS30}, and only one study reports a significantly worse experience in terms of frustration level \cite{PS22}. For instance, \cite{PS27} reports that Copilot reduces both perceived effort and mental demand for novice programmers. Similarly, \cite{PS32} develops a custom questionnaire to assess cognitive load during programming exam tasks. Their findings show that students using Google Bard report lower mental effort compared to those relying on conventional search engines.

In contrast, \cite{PS22} observes no significant difference in overall cognitive load between students using ChatGPT (GPT-4) and those using a traditional web browser, but does report a statistically significant increase in frustration for the ChatGPT group. Similarly, \cite{PS10} finds that participants rate LLM-assisted tasks as equally demanding and effortful as tasks completed without LLM-assistants. Lastly, \cite{PS30} finds no statistically significant differences across all NASA-TLX dimensions when comparing coding with and without ChatGPT (GPT-3.5). 

The variability in reported effects of LLM-assistants on cognitive load highlights the complexity of evaluating mental effort in software development settings. These differences likely stem from diverse operationalizations of cognitive load, differences in participants' expertise, task design, and the capabilities of LLM-assistants across studies. This highlights the need for more standardized methodologies and multi-modal assessment strategies to draw robust conclusions about the cognitive impact of LLM-assistants.


\subsubsection{Econometric analysis}
\label{sec:econometric_analysis}
Productivity is a concept primarily inherited from economics and project management. Two complementary studies investigate the impact of LLM-assistants using quantitative econometric analysis of productivity metrics \cite{PS42, PS46} (see Table \ref{tab:instruments}). The first study \cite{PS42} leverages Time-Cost-Quality (TCQ) conceptual framework to conduct a comparative survey of over 1,000 large firms from 2021 to 2023, examining the effect of GenAI on labor productivity across different domains, including coding and content production. 

Productivity is assessed in terms of both throughput (i.e., time efficiency) and quality (i.e., correctness of output). The study finds that coding exhibits the highest reported gains, with an average 24\% improvement in throughput and 26\% in quality. A complementary study by the same author \cite{PS46} investigates the use of LLM-based pair programming through a survey of 70 large global companies. While the findings confirm that LLM-assistants can enhance development throughput, the study also identifies a critical trade-off: increased throughput is negatively correlated with code quality (r = –0.45). The study suggests that while LLM-assistants can enhance productivity, their effectiveness depends heavily on organizational readiness and the ability to balance speed with software quality.

\begin{rqsummary}{RQ1: Summary}
\DIFadd{Among all primary studies}, laboratory experiments are the most common strategy \DIFadd{(38\%, 15 out of 39)}. Mixed-methods designs are prevalent \DIFadd{(69\%, 27 out of 39)} among empirical methods, often combining user experiments with surveys. 
Time to completion is the most frequently used performance metric \DIFadd{(31\%, 12 out of 39)}. Acceptance rate is a frequently used behavioral metric, though some studies caution against its overuse. Cognitive load findings are mixed: 6 studies use NASA-TLX, but results vary from reduced effort to increased frustration. 
\end{rqsummary}

\section{RQ2: What is the impact of LLM-assistants on software developer productivity?}
\label{sec:RQ2}

\begin{figure}[h]
\centering
\includegraphics[width=1.0\textwidth]{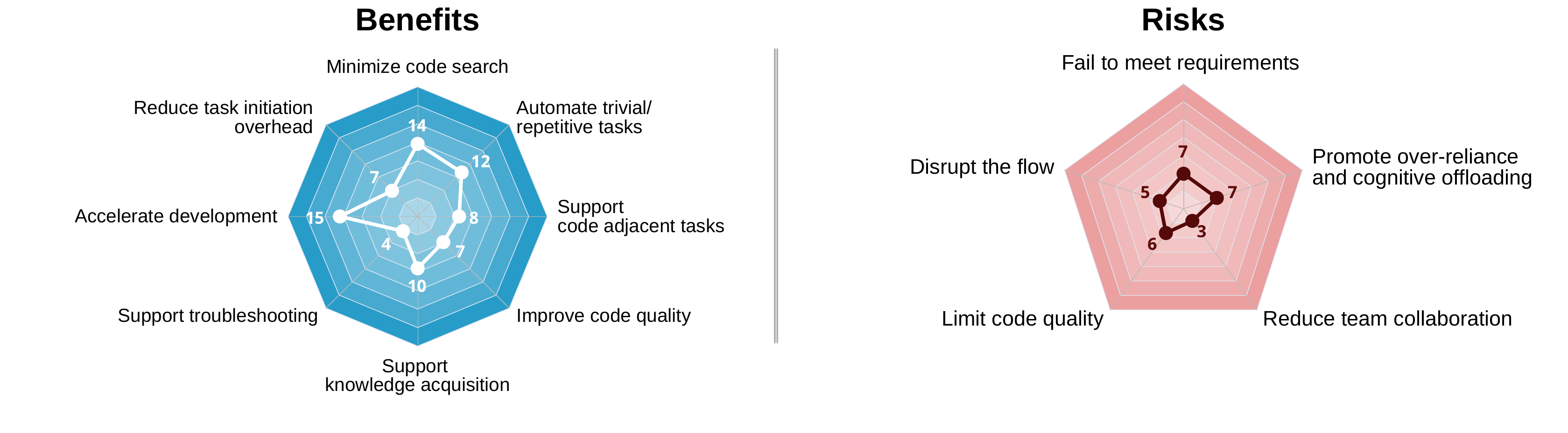}
\caption{Radar plots summarizing how frequently each theme appears as benefit (left) and risk (right) across primary studies on LLM-assisted development.}
\label{fig:SpiderChart}
\end{figure}

We conduct a thematic analysis of the findings reported in each primary study. Our analysis reveals several recurring themes, with multiple studies identifying common benefits and risks associated with the use of LLM-assistants. Figure~\ref{fig:SpiderChart} summarizes the frequency of discovered themes, where each theme is reported as a benefit or risk across the primary studies.

\subsection{Benefits}

\begin{table}[h]
\renewcommand{\arraystretch}{1.5} 

\centering
\caption{Summary of LLM-assistants benefits on developer productivity}
\label{tab:benefits_findings}
\resizebox{\textwidth}{!}{%
\begin{tabular}{|p{2.7cm}|p{12.5cm}|}
\hline
\textbf{Theme} & \textbf{Summary} \\
\hline

Accelerate software development & Studies highlight through self-reported methods that LLM-assistants accelerate software development \DIFadd{\cite{PS1, PS2, PS3, PS13, PS29, PS43, PS59, PS60, PS61}}, and quantitative measures demonstrate that LLM-assistants can reduce task completion time \DIFadd{\cite{PS5, PS20, PS30, PS40, PS57, PS65}}. \\
\hline

Minimize online code search & Study participants in \cite{PS1, PS2, PS3, PS8, PS25, PS29, PS36} noted that LLM-assistants reduce the effort of traditional online search\DIFadd{, with developers preferring them over Stack Overflow and search engines \cite{PS60, PS69}}. Controlled experiments also report benefits over online search \cite{PS30, PS31, PS40} with some studies having mixed results \cite{PS23, PS51}.
\\
\hline

Automate trivial/ repetitive tasks &  LLM-assistants help minimize repetitive coding \cite{PS2, PS11, PS21, PS25} by generating boilerplate code \DIFadd{\cite{PS1, PS2, PS3, PS10, PS43, PS55, PS60}} and reducing keystrokes and typing effort \DIFadd{\cite{PS2, PS8}}. \DIFadd{Test generation and CI/CD automation are key use cases \cite{PS57}.}\\
\hline

Support knowledge acquisition & Studies find LLM-assistants helpful in learning and knowledge acquisition as a direct \DIFadd{\cite{PS2, PS11, PS21, PS25, PS36, PS60, PS69, PS61}} and indirect benefit \cite{PS6, PS10}. LLM-assistants are commonly used as an expert consult, with 75\% of respondents finding them helpful for learning \cite{PS11} \DIFadd{and lowering the entry barrier to a new frameworks \cite{PS60, PS69}. } \\
\hline

Support code-adjacent tasks & LLM-assistants are found helpful in the ideation process \cite{PS36}, requirements specifications \cite{PS5, PS36}, documentation \DIFadd{\cite{PS2, PS3, PS25, PS36}}, and quality assurance \cite{PS2, PS31}. \DIFadd{Developers also use LLM tools for emails, meeting minutes, onboarding documentation, and documenting issues \cite{PS60, PS65}.} \\
\hline

Reduce task initiation overhead & Participants report benefits at the early stages of projects \cite{PS1, PS10} and highlight LLM-assistants' ability to reduce the entry barrier \cite{PS31}. LLM-assistants also support building proof-of-concept applications \cite{PS2}. Developers utilize these tools to generate initial code scaffolding \cite{PS30}\DIFadd{, planning and initial structuring of ideas \cite{PS60, PS61}}.\\
\hline

Improve code quality & LLM-assistants have the ability to enhance the quality of code \cite{PS1}, which is seen as a key advantage \cite{PS25}. \DIFadd{Studies find improvement in code quality \cite{PS51, PS61}}, with metrics such as cyclomatic complexity, code coverage, technical debt, defect density \cite{PS33}, code translation error rate \cite{PS10}, code smells \cite{PS33, PS55}, defect rate \cite{PS33}, and number of defects \cite{PS55}.\\
\hline

Support debugging/ troubleshooting & Participants leverage LLM-assistants to help interpret error messages \cite{PS36} and suggest potential fixes \cite{PS43} without the need to consult extensive documentation \cite{PS25}. \DIFadd{LLM-assistants enable faster bug identification and early defect detection \cite{PS61}.}\\ \hline
\end{tabular}
}
\end{table}








\label{sec:benefits}
\subsubsection{Accelerate software development}
\label{sec:Benefit:accelerateSoftwareDevelopment}

Time remains a central consideration in many definitions of developer productivity \cite{johnson2019effect, jones1994software, zhou2010developer, devanbu1996analytical} as time is both a valuable and constrained resource within development workflows. Participants from several empirical studies, particularly those using self-reported methods, suggest that LLM-assistants can accelerate software development \DIFadd{\cite{PS1, PS2, PS13, PS43, PS59, PS60, PS61}}. Participants often report that LLM-assistants help maintain a state of flow (e.g., ``stay in the flow'') \cite{PS1, PS2, PS29} and contribute to a perceived increase in productivity \DIFadd{\cite{PS1, PS59}}. Supporting this perception, a qualitative analysis of open-ended feedback on an LLM-based code completion tool \cite{PS3} finds ``accelerate coding'' to be the second most frequent theme, mentioned in 14 responses (20\%). \DIFadd{Additionally, a 10-week field study with 90 developers at a large firm reports that 52\% of respondents perceived productivity boosts from using GitHub Copilot, with ratings progressively increasing week-over-week \cite{PS59}.}

Complementing these self-reports, studies using quantitative measures demonstrate that LLM-assistants can reduce task completion time \DIFadd{\cite{PS20, PS30, PS40, PS5, PS57}}. For instance, a case study on the integration of LLMs in the software development lifecycle (SDLC) of a pension plan website \cite{PS5} finds that the required effort decreased from 75 person-days to 22 person-days, representing a productivity gain of 71\%. Similarly, statistical significance in time completion has been observed in coding puzzles. This aligns with one of the modes of human–AI interaction described by \textcite{barke2023grounded} as the acceleration mode. \DIFadd{Controlled experiments also report efficiency gains ranging from 21\% to 45\% depending on task type \cite{PS57}. An analysis of 608 GitHub project teams, comparing matched human-only and human-bot teams based on repository activity metrics, finds that human-bot teams showed significantly higher productivity across all team sizes \cite{PS65}. }

\subsubsection{Minimize online code search}
\label{sec:benefit:OnlineSearch}
Minimizing online code search is the most frequently discovered finding that highlights LLMs benefit (see Figure \ref{fig:SpiderChart}). Indeed, multiple studies highlight the potential of LLM-assistants to reduce the effort required for information retrieval \cite{PS1, PS2, PS3, PS8, PS25, PS29, PS36}, with developers often preferring LLM-assistants over searching for solutions via traditional online resources, including Q\&A platforms such as Stack Overflow \DIFadd{\cite{PS1, PS2, PS8, PS25, PS60, PS69}} and search engines like Google and Bing \cite{PS25, PS69}.

Reducing the need for online search offers several benefits, including helping developers maintain a state of flow (e.g., ``stay in the flow'') \cite{PS29}, improving the speed of syntax recall \cite{PS2}, facilitating the discovery of unfamiliar APIs \cite{PS2, PS3}, and providing an alternative in situations where online search methods fail to deliver \cite{PS25}. LLM-assistants are perceived as a more productive alternative to conventional code search, even when additional effort is required for validation \cite{PS36}.

Several controlled experiments examine this shift by directly comparing LLM-assisted workflows with traditional online search \cite{PS40, PS31, PS23}. In a user study involving 24 participants, \cite{PS31} employs the \textit{SPACE} framework to compare three conditions: traditional web search, code completion (i.e., Copilot), and interaction with a conversational agent (i.e., ChatGPT), finding significant productivity gains for both code completion and conversational agents across all five \textit{SPACE} dimensions—satisfaction, performance, activity, communication, and efficiency. Similarly, \cite{PS40} shows that using an LLM-assistant plugin for code comprehension leads to statistically significant improvements in task completion rates compared to conventional web search. However, \cite{PS23} reports no statistically significant differences in task completion time or correctness when introducing a PyCharm plugin designed to reduce reliance on Stack Overflow, suggesting that benefits vary across tools or tasks.

This task-dependence is further supported by comparative evidence. \cite{PS51} conducts a study with 44 participants comparing ChatGPT and Stack Overflow across algorithmic problems, library usage, and debugging tasks, finding higher-quality outputs for ChatGPT in algorithmic and library tasks, while Stack Overflow performs better for debugging, with no statistically significant difference in task completion time.

\DIFadd{Qualitative interviews contextualize these findings by investigating how developers interpret this trade-off in practice. Several developers report transitioning from Stack Overflow to ChatGPT as a primary information source due to faster, more tailored responses \cite{PS60}.}


\subsubsection{Automate trivial/ repetitive tasks} 
\label{sec:Benefit:Automatetrivial}

Using LLM-assistants help minimize repetitive coding \DIFadd{\cite{PS2, PS11, PS21, PS25}} and reduces trivial tasks by generating boilerplate code \DIFadd{\cite{PS1, PS2, PS3, PS10, PS43, PS55, PS60}}. More specifically, some studies report a reduction in keystrokes and typing effort \DIFadd{\cite{PS2, PS8}}. \DIFadd{Test generation emerges as a key automation use case, with developers viewing unit tests as repetitive tasks well-suited to LLM assistance \cite{PS57}.} The broader impact of offloading cognitively repetitive work is further highlighted through a Delphi judgment study conducted with 14 industry professionals to discuss the future of SE in the age of LLM-assistants \cite{PS21}. The study anticipates that LLM-assistants could be used to automate all routine tasks, hence freeing up developer time for more complex tasks.


\subsubsection{Support knowledge acquisition}

\label{sec:Benefit:knowledgeAcquisition}

The benefits of LLM-assistants extend beyond artifact generation (e.g., source code, test cases). Professional developers increasingly perceive these tools as a valuable aid for learning and knowledge acquisition \DIFadd{\cite{PS2, PS36, PS61}}. LLM-assistants are found to support knowledge acquisition both as a direct \DIFadd{\cite{PS2, PS11, PS21, PS25, PS36, PS60, PS69}} and indirect benefit \cite{PS6, PS10}. The authors of \cite{PS10} highlight the concept of knowledge acquisition as an indirect benefit of using LLM-assistants. Although the main goal is to speed up the development process, 69\% of participants report that the employed code translation LLM-assistant enhanced their learning (e.g., taught them new aspects of Python) \cite{PS10}.

Further evidence is provided by \textcite{PS11}, who analyze developer interactions with ChatGPT. Their findings show that expert consultation is the most common use case, accounting for 62\% of the analyzed conversations \cite{PS11}. Participants from the same study further highlight such benefits, where 75\% of survey respondents report ChatGPT as a helpful learning tool \cite{PS11}. \DIFadd{LLM-assistants are also found to lower the barrier of learning new frameworks, enabling developers to start immediately rather than navigating extensive tutorials \cite{PS60, PS69}. Additionally, one} study involving 14 experts in SE finds that enhancing learning and teaching is the most probable future scenario with LLM-assistants \cite{PS21}. 

\subsubsection{Support code-adjacent tasks}
\label{sec:Benefit:Support-code-adjacent-tasks}

The benefits of LLM-assistants extend beyond coding tasks to support code-related activities. For instance, studies highlight the use of LLM-assistants for ideation \cite{PS36} by exploring different solutions options, requirements specifications \cite{PS5, PS36} including functional and non-functional requirements, documentation \DIFadd{\cite{PS2, PS3, PS25, PS36}} such as in-code documentation and API documentation, and quality assurance \cite{PS2, PS31}. \DIFadd{Developers also use LLM tools for composing emails, generating meeting minutes, and creating onboarding documentation \cite{PS60}. AI assisted teams document significantly more GitHub issues and coordinate more effectively through improved information externalization \cite{PS65}.}



\subsubsection{Reduce task initiation overhead}
\label{sec:Benefit:task_initiation}

A common reported benefit across primary studies is the use of LLM-assistants to support task or project initiation \DIFadd{\cite{PS1, PS2, PS10, PS31, PS30, PS60, PS61}}. Several studies note that developers rely on these tools as a\textit{starting point} for a project or task \cite{PS1, PS10}, effectively lowering the entry barrier \DIFadd{\cite{PS31}}. These tools help developers build momentum by reducing the time and cognitive effort required during the early stages of a project. For example, developers highlight how LLM-assistants can accelerate the development of proof-of-concept applications by generating multiple candidate implementations for the same task \cite{PS2} \DIFadd{generating an initial structure when starting new topics \cite{PS61}.} 

In a qualitative controlled experiment, \cite{PS30} analyzes the interactions of the developers with ChatGPT and finds that 55\% of participants (17 out of 31) used the assistant primarily to generate initial code scaffolding. After this initial phase, participants transitioned to more independent workflows by refining and correcting the code themselves and only relying on the LLM-assistant for targeted questions.



\subsubsection{Improve code quality}
\label{sec:Benefit:Improvecodequality}

Studies highlight the ability of LLM-assistants to improve code quality \DIFadd{\cite{PS1, PS10, PS25, PS33, PS55, PS61}}. Interview participants in \cite{PS1} report using these tools to rewrite and improve the quality of the code. Similarly, 14\% of survey respondents in \cite{PS25} identify improved code quality as a key advantage of LLM-assistants.

These self-reported perceptions are further supported by empirical evidence. \cite{PS33} compares ten projects developed with the support of LLM-assistants and ten developed without such assistance. The authors evaluate six code quality metrics, including cyclomatic complexity, code coverage, code smells, technical debt, and defect density. They report an 18\% improvement across all metrics for projects developed with LLM-assistants. Similarly, \cite{PS55} conducts a case study involving five development teams to examine changes in code quality before and after adopting Copilot. Their findings indicate that three of the five teams experience a measurable reduction in code smells, and all five teams show a decrease in the number of software defects following the integration of LLM-assistants (i.e., Copilot), into their development workflows.

Two controlled experiments further highlight these code quality improvements. In the controlled study \cite{PS51}, the authors compare the code produced by participants using ChatGPT (i.e., treatment group) with the one produced by those using Stack Overflow (i.e., control group). The results show higher code quality for the ChatGPT group in algorithmic and library usage tasks, although the control group outperformed in debugging tasks. Similarly, \cite{PS10} evaluates code translation quality with and without LLM-assistant (i.e., TransCoder). The authors measure error rates using several translation-related metrics (e.g., Translation Error, Language Error, Spurious Error) and find that the group supported with LLM-assistants exhibits fewer translation errors compared to the control group, resulting in a 51\% reduction in error rate.

\subsubsection{Support troubleshooting / debugging}
\label{sec:Benefit:SupportTroubleshooting}

\DIFdel{Despite the use of earlier generations of bots for automating troubleshooting tasks in software development \cite{PS14}, the emergence of LLMs represents the continuation and expansion of this use case.} As illustrated in Figure~\ref{fig:SpiderChart} and Table~\ref{tab:benefits_findings}, recent studies show that LLM-assistants now play an active role in supporting developers during debugging and troubleshooting \cite{PS36}. These tools help interpret error messages by explaining potential causes and suggesting actionable fixes \cite{PS43}. Additionally, several developers report that LLM-assistants accelerate the debugging process, \DIFadd{by enabling faster bug identification and early defect detection by recognizing patterns and errors that developers may miss at manual check \cite{PS61},} and eliminating the need to consult extensive documentation \cite{PS25}.

\subsection{Risks}

\begin{table}[h]
\renewcommand{\arraystretch}{1.5}
\centering
\caption{Summary of reported risks associated with LLM-assistants in developer productivity}
\resizebox{\textwidth}{!}{
\label{tab:challenges}
\begin{tabular}{|p{2.5cm}|p{12.7cm}|}
\hline
\textbf{Theme} & \textbf{Summary} \\
\hline

Fail to meet requirements & LLM-assistants often do not meet functional or non-functional requirements \cite{PS2, PS8}. Developers suggest that not all LLM-assistants' outputs have good accuracy \cite{PS1, PS13}. This is due to the limited controllability of LLMs' output \cite{PS8}, as they can be out of context \cite{PS8, PS43} and tend to over-deliver by providing too much information or repetitive code \cite{PS8, PS31}. \DIFadd{Code generation correctness ranges from 31--65\% \cite{PS57}.}\\
\hline

Promote over-reliance and cognitive offloading & Several studies raise concerns on the
 diminishing of critical thinking skills among novices and students \cite{PS4, PS22, PS47}. In professional settings, instances of automation complacency have also been reported \cite{PS32}. \DIFadd{Developers worry about skill erosion and loss of creativity \cite{PS69}.} To address these issues, studies recommend promoting cautious and informed LLM-assistants use, emphasizing interactive engagement over passive acceptance \DIFadd{\cite{PS30, PS47, PS60}}.\\
\hline

Limit code quality & Concerns were raised about the quality and accuracy of LLM-generated codes \cite{PS25, PS29}. Vulnerabilities and bugs can be introduced if a developer overestimates the capabilities of the tool \cite{PS43}. Working with LLM-assistants might not yield better code quality \cite{PS30, PS46}.\\
\hline

Disrupt the flow & LLM-assistants can disrupt developers' flow with unwanted suggestions \cite{PS1}, interface switching \cite{PS31}, verbose answers \cite{PS31}, and inadequate speed of code suggestions \cite{PS31}. The simultaneous use of multiple code completion tools can disrupt developer flow, particularly when competing suggestions are presented \cite{PS3}. \DIFadd{Developers spend an average of 51.5\% of coding time in LLM interaction states \cite{PS29}. Human-AI teams may experience notification fatigue \cite{PS65}.}\\
\hline

Reduce team collaboration & Relying on LLM-assistants introduces the risk of hindering team collaboration and communication \cite{PS11}. \DIFadd{Traditional help channels have become less active as developers prefer AI assistance \cite{PS69}. Teams report losing organic conversations and synergy \cite{PS69}.} This highlights the need to investigate the impact of LLM-assistance for both human-human and human-agent collaboration and communication \cite{PS37}.\\
\hline

\end{tabular}
}
\end{table}




\label{sec:challenges}

Table~\ref{tab:challenges} shows several risk themes identified across the primary studies that may affect developer productivity. In this section, we describe the five risks categories: limit code quality, fail to meet requirements, promote over-reliance and cognitive offloading, reduce team collaboration, and disrupt the flow.

\subsubsection{Fail to meet requirements}
\label{subsub:accuracy}
Developers acknowledge that not all the suggestions of LLM-assistants are accurate \cite{PS1}. Many survey participants in \cite{PS2, PS8} mention that LLM-assistants often fail to meet both functional and non-functional requirements. Some developers perceive them as difficult to control \cite{PS8}, noting instances where responses are out of context \cite{PS8, PS43} or tend to over-deliver by providing too much information or repetitive code \cite{PS31, PS8}. For example, 50\% of participants report missing or misunderstanding the requirement context as the two main issues encountered with ChatGPT 3.5 \cite{PS43}. Generic or inaccurate code suggestions often require additional effort to modify and refine, leading to an iterative process of prompt refinement and learning how to interact effectively with LLM-assistants \cite{PS13}. \DIFadd{Benchmarking studies quantify these limitations, with code generation correctness scores ranging from 31.1\% (Amazon CodeWhisperer) to 65.2\% (ChatGPT) \cite{PS57}.}



\subsubsection{Promote over-reliance and cognitive offloading}
\label{sec:Risk:over-reliance}

A heavy reliance and excessive trust in LLM-assistants raises concerns about the erosion of critical thinking skills, especially for novice developers and students \cite{PS4, PS22, PS47}. For instance, authors of \cite{PS4} develop an AI tutor that limits direct interaction with ChatGPT through predefined prompts, aiming to promote critical thinking and reduce dependence on LLM-assistants for every minor challenge. However, students still expressed concerns post-experiment, noting that reliance on the LLM tutor might hinder their learning progress \cite{PS4}. The authors acknowledge this issue and highlight the need for further refinement of the tool.

Over-reliance and automation complacency have also been documented among professional software engineers. In a study involving a programming exam with Google Bard, \cite{PS32} observes all three characteristics of automation complacency, as described by \textcite{parasuraman2010complacency}: human monitoring of an automated system, infrequent monitoring, and degraded performance. \DIFadd{Concerns about skill erosion extend across experience levels. Survey respondents express concerns about overreliance, reporting diminished ability to think independently \cite{PS69}.}

These findings highlight the need to promote responsible LLM usage. Several studies advocate for more interactive and reflective engagement with LLM outputs \cite{PS30}, cautioning against blind trust in automated responses \cite{PS47} and encouraging users to understand the tools' capabilities and limitations. \DIFadd{Finding the right balance between leveraging AI support and maintaining developer competence remains an open challenge \cite{PS60}.}

\DIFdel{These findings highlight the need to promote responsible LLM usage. Several studies advocate for more interactive and reflective engagement with LLM outputs \cite{PS30}, cautioning against blind trust in automated responses \cite{PS47} and encouraging users to understand the tools' capabilities and limitations.}

\subsubsection{Disrupt the flow}
\label{subsub:DisturbingTheFlow}
Studies find that LLM-assistants can disrupt developer flow \cite{PS1, PS3}. Issues that impact developer state of flow have been attributed to various kinds of interruptions, including unwanted LLM suggestions \cite{PS1}, interface switching, and verbose answers \cite{PS31}. For example, in a laboratory experiment \cite{PS31}, some professional developers find Copilot distracting, as the speed of code suggestions does not allow sufficient time for code understanding. Distraction also occurs when LLM-assistants work in tandem and compete to display suggestions \cite{PS3}.

Authors of \cite{PS29} investigate developers' interaction with code recommendation systems and their impact on flow by modeling user behavior while using tools such as Copilot. Findings show that developers spend an average of 51.5\% of their coding session time in LLM interaction states, such as verifying suggestions, prompt crafting, and deferring thought. \DIFadd{In open source contexts, human-bot teams may experience notification fatigue from increased automated activity \cite{PS65}.} These findings highlight the temporal and cognitive costs that these tools may introduce.



\subsubsection{Limit code quality}
\label{sec:Risk:LimitCodeQuality} \noindent

\DIFdel{Code quality is the one theme that has been reported as both a benefit and a risk (see Figure \ref{fig:SpiderChart}, Table ~\ref{tab:benefits_findings} and ~\ref{tab:challenges}).}

Multiple studies raise concerns about the quality of generated code. For instance, \cite{PS25} reports that 13\% of survey respondents raise concerns about the quality and accuracy of LLM-generated code. Similar concerns are raised by 29\% of survey participants when using Copilot \cite{PS29}. Code quality issues arise when developers overestimate the capabilities of such tools, which can introduce vulnerabilities and bugs. LLM-assistants often struggle with optimization and refactoring tasks, especially when lacking semantic context. Moreover, in \cite{PS43}, 38\% of the surveyed participants point to erroneous code as one of the limitations of ChatGPT, while 29\% report limitations due to inefficient code. User experiments conducted by the authors of \cite{PS30} reveal no significant improvements in code quality or correctness with LLM-assistants, with a slightly worse quality average for the ChatGPT group.

When analyzing the relationship between reported productivity and code quality gains in the context of LLM-assistant, a large industry study involving 70 large global companies \cite{PS46} reports a moderate negative correlation ($r=-0.45$) between the two. These findings highlight that increased productivity through the support of LLM-assistants does not necessarily lead to improvements in code quality. \DIFadd{Hallucinations remain a persistent concern \cite{PS60}. LLM-assistants may produce non-functional solutions or outputs containing deprecated libraries that fail review \cite{PS60}.} \DIFadd{The lack of contextual understanding extends to organizational and domain-specific knowledge. LLM-assistants struggle to account for company-specific coding standards, architectural conventions, or project constraints, potentially generating code that violates established conventions of the surrounding codebase \cite{PS60}.}

Code quality is the one theme that has been reported as both a benefit and a risk (see Figure \ref{fig:SpiderChart}, Table ~\ref{tab:benefits_findings} and ~\ref{tab:challenges}). The variation in findings across studies can be attributed to differences in both experimental scope and evaluation methods. Some studies examine small, isolated programming tasks, while others assess complex, real-world systems. Moreover, researchers use diverse metrics (e.g., such as cyclomatic complexity, defect density, and code coverage) that capture different dimensions of code quality. These contextual and methodological differences make direct comparison difficult and help explain why LLM-assisted code quality appears both improved and degraded across the literature.


\subsubsection{Reduce team collaboration}
\label{subsub:teamcollaboration}
Relying on LLM-assistants can negatively impact productivity by reducing team collaboration and communication \cite{PS11}. For instance, a field study \cite{PS11} observes that excessive use of LLM-assistants may lead developers to favor consulting a chatbot over a colleague. In fact, the overconfidence of LLM-assistants' responses can create the impression that team discussions are unnecessary, reducing opportunities for communicative learning and discovery \cite{PS11}. The intrinsic nature of LLM-assistants also plays a role in reduced team collaborations. For instance, current conversational LLM-assistants offer limited support for team collaboration, as they primarily support one-on-one interactions and are not well-suited to facilitating effective team coordination.
\DIFadd{Traditional help channels have become less active, with developers preferring AI assistance, in part to avoid exposing knowledge gaps to colleagues \cite{PS69}. Teams report losing the organic conversations and synergy that come from bouncing ideas off each other \cite{PS69}.} These findings highlight the need for future studies to further investigate how LLM-assistants affect human-human collaboration and how they can be designed to foster team collaboration \cite{PS37}.


\begin{rqsummary}{RQ 2 - Summary}
Studies report mixed findings on the use of LLM-assistants, revealing both notable benefits and critical risks. The most frequently reported benefits include \DIFadd{accelerated development, minimizing code search, and} automating trivial or repetitive tasks. At the same time, primary studies identify several risks, such as \DIFadd{failing to meet requirements,} promoting over-reliance and cognitive offloading, and \DIFadd{disrupting developer flow}. Code quality emerges as a particularly contested area, with evidence pointing to both improvements and degradations depending on context. These discrepancies underscore the need for further investigation and the development of strategies to ensure that code quality is maintained when integrating LLM-assistants into software development workflows.
\end{rqsummary}

\section{RQ3: Which dimensions of developer productivity are investigated, and how do these dimensions map onto the SPACE framework?}
\label{sec:RQ3}

The diverse methodologies and wide variation in reported findings across primary studies (see Section~\ref{sec:RQ1} and Section~\ref{sec:RQ2}) highlight the need for a structured lens to interpret how productivity is conceptualized in the literature, especially in the context of rapidly evolving development with LLM-assistants. However, existing approaches to measuring productivity often fall short of capturing this diversity. Traditional metrics such as Lines of Code (LOC) and Function Points, have long been criticized for offering only a narrow view of productivity centered on output quantity. Likewise, quantitative performance frameworks such as the DevOps Research and Assessment (DORA) model \cite{google2022devops}, though valuable for assessing delivery speed and deployment frequency, do not reflect the human, social, and cognitive dimensions that shape productivity in Human–AI Interaction. Recognizing these limitations, we adopt a multidimensional perspective grounded in the \textit{SPACE} framework, which integrates objective outcomes (e.g., performance, activity, efficiency) with human-centered constructs (e.g., satisfaction and collaboration).

\begin{figure}[h]
    \centering
    \includegraphics[width=0.85\textwidth]{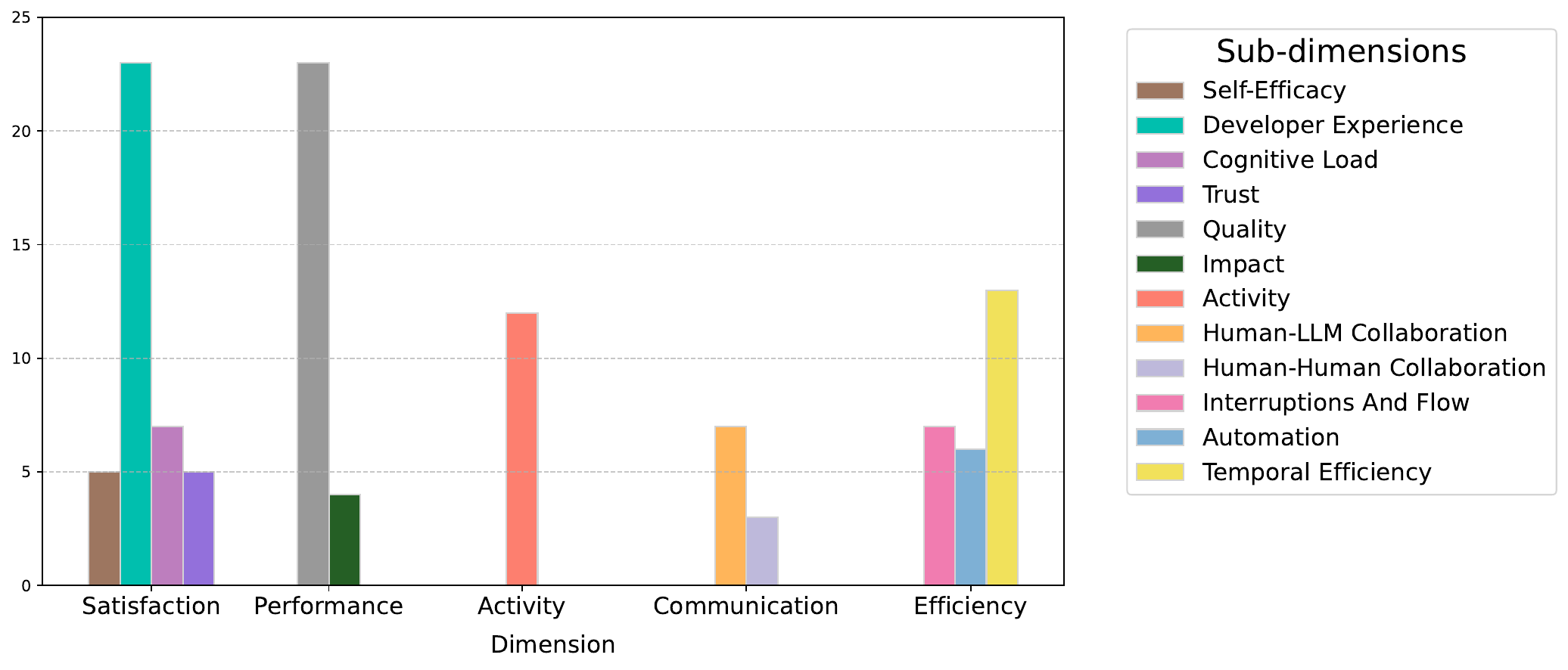}
    \caption{Distribution of sub-dimensions across each dimension of the \textit{SPACE} framework.}
    \label{fig:subcategory_frequency}
\end{figure}

Therefore, we leverage the \textit{SPACE} framework \cite{forsgren2021space}, a widely used framework developed by Microsoft researchers, as an organizing model to understand productivity from a practical perspective. We choose the \textit{SPACE} framework because it defines productivity as a multi-dimensional construct that covers a diverse range of study instrumentation from measurement metrics (e.g., code quality) to perceptions (e.g., satisfaction). The \textit{SPACE} framework also reflects a modern development workflow, aligned with the complexities of LLM-assisted development, which emphasizes dimensions like collaboration and trust. Its extensible and adaptable nature allows for contextual adaptation, as it does not prescribe fixed metrics but offers a conceptual structure that can be tailored to specific empirical contexts. We map each primary study to the five dimensions of the \textit{SPACE} framework: Satisfaction, Performance, Activity, Communication \& collaboration, and  Efficiency (see Figure~\ref{fig:subcategory_frequency}).

\begin{table}[h]
\centering
\caption{Mapping of primary studies to SPACE dimensions and derived sub-dimensions. With sources for each sub-dimension are cited from prior work.}
\resizebox{\textwidth}{!}{%
\begin{tabular}{|l|l|p{7cm}|c|}
\hline
\textbf{Dimension} & \textbf{Sub-dimensions} & \textbf{Primary Studies} & \textbf{\%} \\
\hline
\multirow{5}{*}{Satisfaction} 
 & Developer experience \cite{razzaq2024systematic} 
 & \cite{PS2, PS3, PS4, PS8, PS9, PS11, PS13, PS20, PS23, PS25, PS28, PS29, PS31, PS33, PS36, PS40, PS43, PS47, PS51, PS55, PS59, PS60, PS61} & \multirow{5}{*}{\DIFadd{77\%}} \\
\cline{2-3}
 & Self-efficacy \cite{forsgren2021space, sikand2024much} 
 & \cite{PS1, PS22, PS27, PS37, PS61} & \\
\cline{2-3}
 & Trust \cite{sikand2024much} 
 & \cite{PS11, PS22, PS32, PS37, PS43} & \\
\cline{2-3}
 & Cognitive load 
 & \cite{PS10, PS20, PS22, PS27, PS30, PS32, PS40} & \\
\hline
\multirow{2}{*}{Performance} 
 & Quality \cite{forsgren2021space, sikand2024much} 
 & \cite{PS3, PS5, PS10, PS20, PS22, PS23, PS27, PS28, PS29, PS30, PS31, PS32, PS33, PS37, PS42, PS43, PS46, PS47, PS48, PS51, PS55, PS57, PS69} & \multirow{2}{*}{\DIFadd{64\%}} \\
\cline{2-3}
 & Impact \cite{forsgren2021space} 
 & \DIFadd{\cite{PS21, PS42, PS46, PS60}} & \\
\hline
Activity &  
 & \cite{PS3, PS6, PS9, PS22, PS28, PS29, PS31, PS47, PS48, PS55, PS65, PS69} & \DIFadd{31\%} \\
\hline
\multirow{2}{*}{Communication} 
 & Human-LLM collaboration 
 & \cite{PS4, PS28, PS29, PS31, PS36, PS40, PS60} & \multirow{2}{*}{\DIFadd{26\%}} \\
\cline{2-3}
 & Human-human collaboration 
 & \cite{PS11, PS37, PS55} & \\
\hline
\multirow{3}{*}{Efficiency} 
 & Temporal efficiency \cite{forsgren2021space} 
 & \cite{PS5, PS6, PS23, PS30, PS31, PS32, PS33, PS40, PS46, PS51, PS55, PS57, PS69} & \multirow{3}{*}{\DIFadd{59\%}} \\
\cline{2-3}
 & Interruptions and flow \cite{forsgren2021space, sikand2024much} 
 & \cite{PS1, PS2, PS13, PS21, PS28, PS29, PS37} & \\
\cline{2-3}
 & Automation 
 & \cite{PS2, PS9, PS25, PS29, PS57, PS60} & \\
\hline
\end{tabular}
}
\label{tab:space-table}
\end{table}

\begin{figure}[h]
    \centering
    \includegraphics[width=0.85\textwidth]{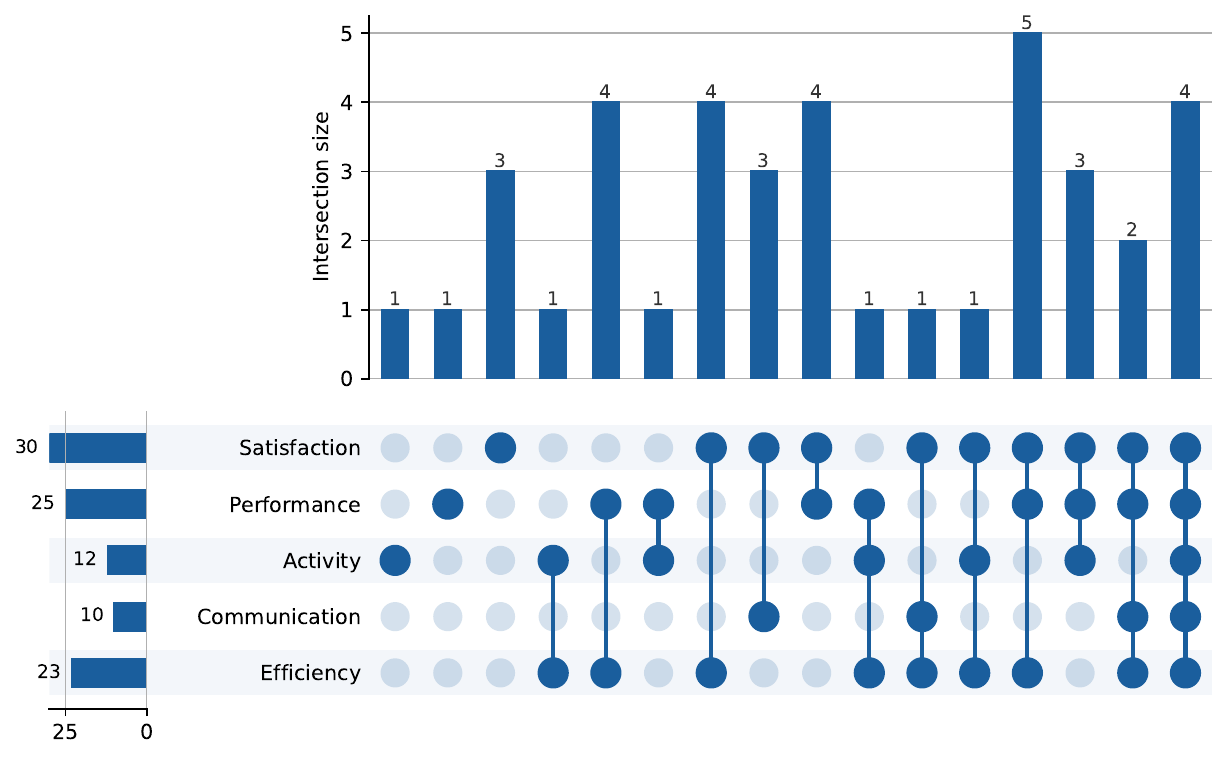}
    \caption{The distribution of investigated SPACE dimensions and their overlap.}
    \label{fig:space_upset}
\end{figure}

To synthesize and compare findings from our primary studies, we leverage thematic analysis, employing an adapted version of the \textit{SPACE} framework. We use the five dimensions of the \textit{SPACE} framework. To provide more granularity, we further refine these dimensions by including sub-dimensions. Specifically, we adapt some sub-dimensions from relevant related work \cite{sikand2024much}. When a specific concept was not captured by an existing sub-dimension, we ensure comprehensive coverage by including emerging sub-dimensions derived from our data.
A structured taxonomy for our analysis is provided in Table~\ref{tab:space-table}. Figure \ref{fig:subcategory_frequency}  illustrates the distribution of the sub-dimensions across each dimension of the \textit{SPACE} framework.


Our findings reveal that the majority of studies \DIFadd{(90\%, 35 out of 39)} adopt a multidimensional perspective on productivity, with only four studies taking a uni-dimensional perspective. This indicates a shift away from singular-dimension perspectives toward a more holistic understanding of the complex nature of productivity in software engineering (See figure ~\ref{fig:space_upset}). However, only \DIFadd{44\%} of the studies \DIFadd{(17 out of 39)} examine three or more of the \textit{SPACE} dimensions, with just \DIFadd{15\% (6 out of 39) addressing four or more dimensions}. This highlights the need for future work to capture the full breadth of how LLM-assistants impact productivity. We find that the most co-occurring combinations involve Satisfaction, Performance, and Efficiency. \DIFadd{Out of the list of primary studies, the most frequent combination is Satisfaction-Performance-Efficiency (5 out of 39)}.

Table ~\ref{tab:space-table} and Figure ~\ref{fig:space_upset} show that \textbf{Satisfaction} is the most studied dimension, addressed by \DIFadd{77\% (30 out of 39)} of primary studies. This dimension captures developers' feelings about their work with LLM-assistants, which is mainly captured through self-reported instruments. Our analysis of this dimension reveals five fine-grained sub-dimensions (i.e., \textit{developer experience}, \textit{self-efficacy}, \textit{trust}, and \textit{cognitive load}). Most studies within the satisfaction dimension focus on the concept of \textit{developer experience}, which encompasses developers' perceptions, feelings, and values regarding their interactions with LLM-assistants \cite{razzaq2024systematic}, as well as the perceived importance of LLM-assistants and their ease of use (e.g., developers' feedback or the Technology Acceptance Model (TAM)). \textit{Cognitive load} is the second most commonly investigated satisfaction sub-dimension, assessed using instruments such as NASA-TLX or custom surveys. \textit{Self-efficacy}, the belief in one’s ability to complete tasks, is explored using both validated and custom-designed tools. Finally, \DIFadd{We find that \textit{well-being} is not examined by any of the empirical studies. This aligns with the observations of \cite{qiu2025today}, who highlight that developers’ well-being and mental health are frequently overlooked in software engineering research.}

\textbf{Performance} \DIFadd{is the second most studied dimension covered by 64\% (25 out of 39) of primary studies}. Performance mainly concerns the final outcomes of software development activities. The majority of studies investigate the \textit{quality} sub-dimension of performance using instruments such as passing unit tests, functional correctness, and code smells (see Table~\ref{tab:performance-metrics}). The \textit{impact} sub-dimension is addressed in only three studies and mainly investigates how LLM-assistants impact final product outcomes \cite{forsgren2021space}. This sub-dimension focuses on business-related metrics, including cost savings, product quality improvements, and delivery speed (see section ~\ref{sec:econometric_analysis}).

\textbf{Efficiency} \DIFadd{is the third most studied dimension covered by \DIFadd{59\% (23 out of 39)} of primary studies}. This dimension reflects the capacity to complete tasks efficiently with minimal interruptions or time delays, as it aims to minimize unnecessary delays and optimize the flow of task handoffs \cite{forsgren2021space}. We highlight three sub-dimensions explored by primary studies (i.e., \textit{temporal efficiency}, \textit{automation}, \textit{interruptions and flow}). Efficiency is often investigated from the temporal perspective, measured using task completion metrics or via developer perceptions. \textit{Automation} is another important angle of efficiency, with studies reporting how LLM-assistants are used to offload repetitive tasks such as writing boilerplate code \cite{PS2, PS25}. Finally, studies examine efficiency in terms of \textit{interruptions and flow}, with some highlighting reduced cognitive interruptions and others noting new forms of distraction introduced by the LLM-assistant.

The \textbf{Activity} \DIFadd{is one of the least explored dimensions (31\%, 12 out of 39).} Activity is often paired with efficiency and performance and focuses on counts and frequency measures while performing a given task \cite{forsgren2021space}. Studies included in our \DIFadd{review} often measure activity as the count of actions or tasks developers take during their interactions with LLMs (e.g., acceptance rate, number of tasks completed)\cite{PS28, PS31, PS3, PS6}. For example, \textcite{PS28} measures the dimension of activity at a finer granularity, regarding the count of actions developers take during their interactions with Copilot (e.g., acceptance rate, suggestions shown rate, completions changed or unchanged).



\textbf{Communication} \DIFadd{is the least investigated dimension across primary studies (26\%, 10 out of 39)}. Communication focuses on how developers and teams communicate and share knowledge \cite{forsgren2021space}. \DIFadd{The majority of the studies mapped to the communication dimension investigate the dimension in terms of \textit{human-LLM collaboration} (7 out of 10)}  , which includes analysis of interaction patterns between participants and LLM-assistants. While only three \DIFadd{(3 out of 10)} examine \textit{human-human collaboration} with LLM in the loop. \DIFadd{This shows a gap in our understanding of how LLM-assistants influence team communication or coordination, which is also highlighted by prior studies \cite{qiu2025today, abrahao2025software}.} Given that emerging concerns regarding reduced team collaboration due to over-reliance on LLM-assistants (see Section \ref{subsub:teamcollaboration}), future studies should incorporate more investigation on team dynamics to better understand their impact in the LLM-assisted workflows.

\begin{table}[h]
\renewcommand{\arraystretch}{1}
\centering
\caption{Quality metrics by study.}
\label{tab:performance-metrics}
\begin{tabular}{|p{5cm}|p{4cm}|}
\hline
\textbf{Metric} & \textbf{Primary Studies} \\
\hline
Passing Unit Tests & \cite{PS10, PS27, PS30, PS43, PS51} \\
\hline
Functional Correctness and Accuracy & \cite{PS5, PS37, PS48, PS51} \\
\hline
Code Smells & \cite{PS22, PS33, PS55} \\
\hline
BLEU Score & \cite{PS3, PS48} \\
\hline
Halstead Complexity Measures & \cite{PS5, PS31} \\
\hline
Cyclomatic Complexity & \cite{PS23, PS33} \\
\hline
Translation Error Rate & \cite{PS10} \\
\hline
Maintainability Index & \cite{PS31} \\
\hline
Cognitive Complexity & \cite{PS55} \\
\hline
Defect Density & \cite{PS33} \\
\hline
Defect Rate & \cite{PS55} \\
\hline
Technical Debt & \cite{PS33} \\
\hline
Code Coverage & \cite{PS33} \\
\hline
\end{tabular}
\end{table}

\begin{rqsummary}{RQ3 - Summary}
Our analysis, framed by the \textit{SPACE} framework, reveals that the majority of studies \DIFadd{(90\%, 35 out of 39)} adopt a multidimensional view of productivity in SE, mostly combining two or more \textit{SPACE} dimensions. However, only \DIFadd{15\% (6 out of 39)} of the studies examine four or more \textit{SPACE} dimensions. \DIFadd{Satisfaction (77\%, 30 out of 39)} is the most frequently investigated, followed by \DIFadd{Performance (64\%, 25 out of 39)} and \DIFadd{Efficiency (59\%, 23 out of 39)}. In contrast, \DIFadd{Activity (31\%, 12 out of 39)} and \DIFadd{Communication (26\%, 10 out of 39)} is the least explored dimensions.
\end{rqsummary}

\section{Discussion}
\label{sec:Discussion}

\DIFadd{This discussion synthesizes findings from 39 primary studies to derive cross-cutting insights, actionable guidance, and research directions on the use of LLM-assistants in software development.
We first present an in-depth synthesis of the findings using McLuhan’s Tetrad~\cite{mcluhan1977laws} to explain how LLM-assistants reshape development practices beyond measurable productivity gains. We then distill key takeaways and lessons learned for practitioners, followed by actionable recommendations at the individual, team, and organizational levels. Finally, we identify open issues and research gaps and discuss their practical and ethical implications for the software engineering community.}

\subsection{In-depth synthesis across studies using the McLuhan Tetrad}  

As established in RQ3 (Section~\ref{sec:RQ3}), traditional approaches to defining and measuring productivity in software engineering remain limited, even when expanded through multidimensional frameworks such as \textit{SPACE}. While these models clarify how productivity is operationalized, they do not fully capture the deeper transformations that arise when new technologies reshape everyday development practices. The integration of LLM-assistants extends beyond measurable performance gains, since it redefines development practices, decision-making, and team dynamics across the entire software life cycle \cite{PS5, PS11, PS55}. As discussed in Section~\ref{sec:benefits}, LLM-assistants bring clear productivity benefits, such as improved code quality and accelerated development, yet also introduce risks related to over-reliance, erosion of critical judgment, and reduced collaboration (Section~\ref{sec:challenges}).

\begin{figure}
    \centering
    \includegraphics[width=0.75\linewidth]{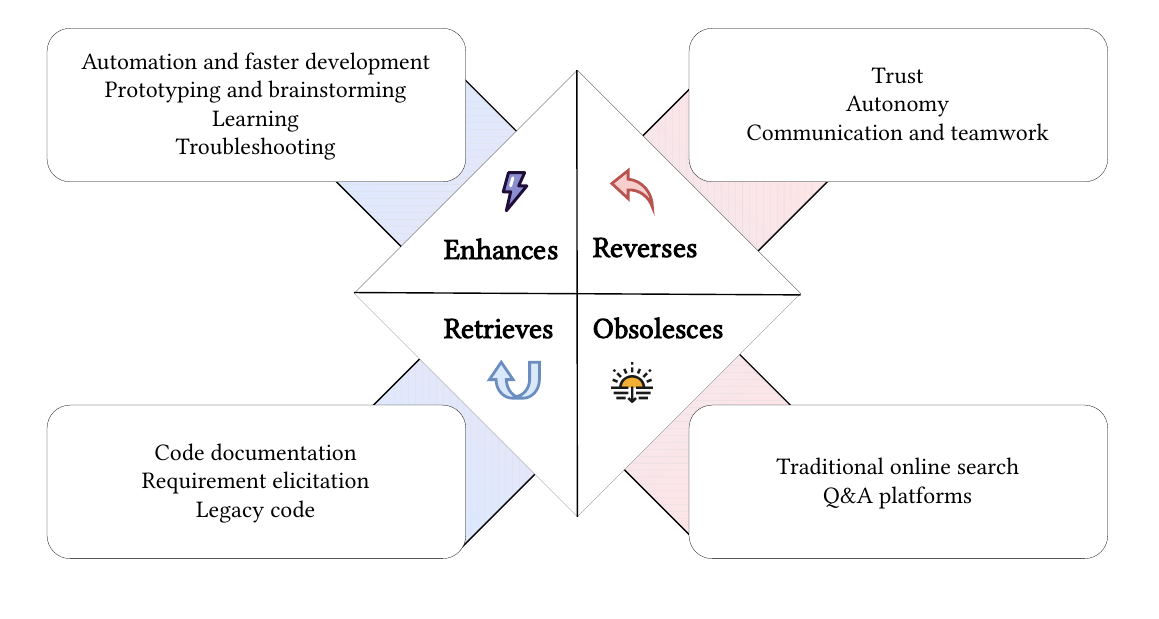}
    \caption{
    McLuhan's Tetrad diagram illustrates the implications of LLM assistants on the productivity of software developers. The diagram captures four dimensions.
    \textit{Enhancement}: how LLM-assistants amplify development speed; 
    \textit{Obsolescence}: which traditional practices are being displaced; 
    \textit{Retrieval}: which previously diminished practices are being revived; and 
    \textit{Reversal}: what adverse effects may emerge when LLM-assistants are pushed to the extreme.
    }
    \label{fig:Tetrad}
\end{figure}

To synthesize these broader transformations, we draw on McLuhan’s tetrad \cite{mcluhan1977laws}, a conceptual framework originally proposed to analyze how emerging media technologies transform human behavior and social organization. The Tetrad complements the \textit{SPACE} framework by shifting the focus from measurement to interpretation. It examines \textit{with what consequences} these socio-technical shifts occur. The framework poses four interrelated questions: 1) \textit{What does the medium enhance?}, 2) \textit{What does it render obsolete?}, 3) \textit{What does it retrieve?}, and 4) \textit{What does it reverse when pushed to the extreme?} Figure~\ref{fig:Tetrad} illustrates how this lens is applied to analyze the impact of LLM-assistants on software development.

\subsubsection{Enhance.} In McLuhan's Tetrad, the \textit{Enhance} dimension refers to what technology intensifies in existing practices. In the realm of SE, LLM-assistants have the potential to enhance several aspects of the development process. As summarized in Section~\ref{sec:benefits}, multiple studies report that LLM-assistants enhance productivity primarily by accelerating development (Section \ref{sec:Benefit:accelerateSoftwareDevelopment}), lowering entry barriers for complex tasks (Section \ref{sec:Benefit:task_initiation}), support knowledge acquisition for both professionals and students (Section \ref{sec:Benefit:knowledgeAcquisition}), and support debugging and troubleshooting (Section \ref{sec:Benefit:SupportTroubleshooting}). \DIFadd{Taken together, LLM assistants are effective when applied to tasks that align with their strengths, such as boilerplate generation, syntax recall, initial scaffolding, and exploratory prototyping. \textbf{Practitioners should leverage LLM-assistants selectively and strategically rather than expecting uniform gains across all development activities.}}

\subsubsection{Reverse.} According to McLuhan's Tetrad, the \textit{Reverse} dimension captures the unintended or counterproductive consequences that arise when technology is pushed to its limits. As discussed in Section~\ref{sec:Risk:over-reliance}, excessive trust in LLM-assistants can lead to cognitive offloading and automation complacency, while the lack of trust creates frustration and discourages adoption. \DIFadd{Over-reliance further contributes to diminished developer autonomy \cite{abrahao2025software} by shifting developers from active code production to reviewing generated output. This reduced reflective engagement can negatively affect} code quality (Section \ref{sec:Risk:LimitCodeQuality}) particularly when LLM-generated code is accepted without sufficient validation. Finally, reliance on LLM-assistants may weaken collaboration and peer communication (Section \ref{subsub:teamcollaboration}), as developers turn to the LLM-assistant instead of teammates for feedback. \DIFadd{Taken together, these findings suggest that productivity gains can diminish when LLM-assistants are used without adequate oversight. Uncritical reliance can undermine reflective practice, collaborative problem-solving, and the development of domain expertise that are essential for long-term software quality. \textbf{Practitioners should therefore use LLM-assistants critically: treat generated outputs as preliminary drafts requiring thorough review, maintain active engagement in coding decisions, and balance tool use with collaborative practices to preserve reflective judgment, team communication, and long-term code quality.}}

\subsubsection{Obsolesce.}
The \textit{Obsolesce} dimension of McLuhan's Tetrad refers to the tools, practices, or roles that are diminished or displaced as a result of adopting new technologies. For LLM-assistants, traditional online search techniques for recalling syntax, or looking up new libraries online (Section \ref{sec:benefit:OnlineSearch}). Developers may prefer using conversational agents over searching for solutions or consulting Q\&A platforms such as Stack Overflow (Section \ref{sec:benefit:OnlineSearch}). \DIFadd{While these effects may enhance efficiency, they also risk diminishing the capacity to independently verify information, evaluate competing solutions, and cultivate the search and validation that remain essential when LLM outputs are incorrect or incomplete. \textbf{Practitioners should therefore treat LLM-assistants as a complement rather than a replacement for traditional information-seeking. They should continue maintaining proficiency in search and validation practices, cross-check suggestions against reliable sources.}}

\subsubsection{Retrieve.} In McLuhan's Tetrad, the \textit{Retrieve} dimension reflects the resurgence or reintegration of practices that had diminished in relevance prior to the adoption of new technology. For LLM-assistants, previously neglected activities such practice code documentation, which is now being brought back to focus in development workflows with the help of LLM-assistants (Section \ref{sec:Benefit:Support-code-adjacent-tasks}), and requirements engineering, including requirement elicitation that are traditionally time-consuming due to frequent client communication \cite{fontanet2014developers}, are gaining momentum through LLM-assistants.

Additionally, LLM-assistants \DIFadd{also offer opportunities} to revisit legacy systems, a domain often overlooked due to the high cost and complexity \DIFadd{and maintaining outdated platforms such as COBOL or Uniface\cite{PS13, PS21}.  While support for legacy systems is currently limited\cite{PS8}, these tools show promise in expanding the scope of productivity beyond coding speed to include documentation, requirements, and modernization tasks. \textbf{Practitioners should leverage LLM-assistants to revive these previously deprioritized practices. Specifically, they can integrate LLM-assisted documentation, requirements elicitation, and legacy system support into development workflows to enhance maintainability, preserve institutional knowledge, and improve long-term software quality, while still combining AI assistance with human expertise.}}

\begin{mygraybox}
\DIFadd{\textbf{\textit{Lessons learned.}} Across enhancement, reversal, obsolescence, and retrieval, three overarching lessons emerge:\\
(1) Productivity gains are task-contingent, strongest for well-scoped and repetitive activities.\\
(2) Uncritical reliance introduces diminishing returns through validation overhead and erosion of reflective practice.\\
(3) LLM-assistants reshape, not replace, developer expertise, shifting effort toward evaluation, judgment, and coordination.}
\end{mygraybox}

\subsection{Implications and recommendations for software developers}

The integration of LLM-assistants into the software development workflow has profound implications, transforming both the individual developer's role and the team's dynamics. While these tools offer benefits, developers must be mindful of the potential risks and adapt their practices to maximize productivity and maintain high-quality code.


\textbf{Cultivating and managing trust.} Establishing the appropriate level of trust, presents challenges caused by the trade-off between benefits and risks, as presented in RQ2 (see Section \ref{sec:RQ2}). In RQ3 (Section \ref{sec:RQ3}), we incorporated trust as a sub-dimension of ``Satisfaction'', since trust can influence how developers perceive their productivity \cite{PS43}. Excessive trust can lead to automation complacency and over-reliance \cite{PS32}, while insufficient trust can cause frustration and underutilization of the tools' benefits. Balancing trust is therefore essential to sustain long-term satisfaction and effective collaboration between developers and LLM-assistants.


Trust can become fragile when these models hallucinate or produce incorrect outputs, such as suggesting non-existent APIs or generating syntactically incorrect code \cite{PS3}. These issues stem from the traditional trade-off between precision and recall in machine learning systems, where models tuned for high recall may increase coverage but at the cost of higher volume of incorrect suggestions \cite{PS3}. This creates a trust barrier between developers and LLM-assistants. 

Additionally, LLMs often suffer from a lack of transparency, as they do not provide sources or references for their outputs \cite{PS11} which deepens the trust gap. Implementing AI solutions with transparency mechanisms \cite{PS25} to elucidate their decision-making processes empowers developers to trust and better understand the LLM-assistants' suggestions and decisions. These implications may extend to a much broader scope. Through a questionnaire conducted by \cite{PS43}, the authors find a moderate positive correlation between self-reported productivity and trust levels, as participants who report increased productivity with LLM-assistants also exhibit slightly higher levels of trust. However, trust dynamics are not uniform across user groups. For instance, the study in \cite{PS32} finds that novice developers tend to demonstrate automation complacency, often accepting LLM-assistants suggestions uncritically and with minimal validation. While this may temporarily boost productivity, it also raises concerns about long-term skill development and critical thinking.

\begin{mygraybox}
\textbf{\textit{Recommendation 1.}}
Developers should cultivate calibrated trust by understanding the capabilities and limitations of LLM-assistants. \DIFadd{Operationally, this recommends: (i) treating all LLM-generated code as preliminary output requiring validation through testing and review; (ii) cross-referencing suggestions against official documentation for unfamiliar APIs; (iii) acknowledging that LLMs lack project-specific context and may produce hallucinated outputs; and (iv) periodically engaging in unassisted coding to preserve fundamental competencies. Cultivating such informed, critical trust is essential for maximizing benefits while safeguarding code quality, developer autonomy, and long-term skill development.}
\end{mygraybox}


\textbf{Redefining the developer's role from coder to reviewer.}
 LLM-assistants often lack awareness of the broader context and intricacies of a complex software project \cite{PS2, PS8}. Developers often mitigate this issue by breaking down the problem \cite{PS2, PS8, PS43} and providing a clear explanation to the LLM-assistants \cite{PS2, PS8}. Consequently, developers increasingly spend time verifying, editing or refining LLM-assistants generated suggestions rather than writing code. One study by \textcite{PS29} reports that participants spend over 50\% of their time in evaluation activities, such as crafting prompts, reviewing suggestions, and editing completion. Similarly, \textcite{PS10} finds that using LLM-assistants for code translation transforms the developer’s responsibility into one of reviewing rather than writing code. Time saved in code generation can be lost in the evaluation and refinement phases, especially for complex tasks. This can lead to diminishing returns on productivity if not managed effectively. \cite{PS46}. These findings align with related work regarding ``ironies of automation'' \cite{simkute2025ironies}, where the introduction of automation often shifts users’ roles from production to evaluation. 
This shift raises practical implications for developers’ evolving roles. New skills such as prompt engineering, critical evaluation of AI-generated code, and contextual judgment are becoming central to day-to-day work. Awareness of automation bias and complacency must also be cultivated to prevent uncritical acceptance of LLM suggestions. Accordingly, developer training should emphasize verification, debugging, reflective use of AI, and collaborative practices that preserve knowledge sharing. At the team level, this role change may alter composition and responsibilities, increasing the value of roles focused on oversight, integration, and code review rather than routine implementation.

\begin{mygraybox}
\textbf{\textit{Recommendation 2.}}
The role of developers is evolving from coder to reviewer when using LLM-assistants. To maximize productivity and code quality, developers must shift their mindset and reallocate time and cognitive resources to tasks like prompt engineering, iterative evaluation, and refining LLM-generated outputs. This includes guiding LLMs with precise context, treating suggestions as drafts requiring human validation, and decomposing complex tasks into smaller, well-scoped sub-problems for more coherent responses.
\DIFadd{Given that evaluation activities may constitute a considerable amount of time, developers are advised to allocate dedicated time for validation rather than treating it as ancillary, and to adopt systematic approaches for assessing LLM outputs with respect to correctness, security, and adherence to project standards.}
\end{mygraybox}

\textbf{Adapting the development workflow.}
The integration of LLM-assistants may disrupt established software development workflow, which can affect team collaboration and individual tasks. At the team level, we describe in section~\ref{subsub:teamcollaboration} how LLM-assistants may reduce collaboration among software teams. It remains unclear to what extent the adoption of LLM-assistants leads to a decline in developer knowledge sharing and pair programming. At the individual level, LLM-assistants can disrupt developers’ flow. As described in section~\ref{subsub:DisturbingTheFlow}, unwanted or irrelevant LLM-assistants' suggestions can disrupt the developer's flow. For instance, studies report that interruptions from tools like Copilot, especially when suggestions are too frequent or conflict with other tools, can disrupt focus and reduce productivity \cite{PS1, PS3, PS31}.

\begin{mygraybox}
\textbf{\textit{Recommendation 3.}}
Developers must adapt their practices and team dynamics to ensure LLM-assistants enhance, rather than hinder, the development workflow. Individually, developers should customize tool settings to control suggestion frequency and limit interruptions. At the team level, preserving collaborative practices like pair programming, code reviews, and architectural discussions is crucial to maintaining communication and knowledge sharing. \DIFadd{Teams are encouraged to establish explicit norms governing when to consult colleagues versus LLM-assistants, particularly for decisions affecting shared codebases, and to document LLM-assisted decisions in commit messages or design records to preserve collective awareness.}
\end{mygraybox}


\textbf{Organizational factors and adoption strategy.} The effectiveness of LLM-assistants depends on organizational context. Studies examining productivity at the organizational level (Section~\ref{sec:econometric_analysis}) analyze how firm-level or team-level practices shape the adoption and outcomes of LLM-assisted development. Evidence from industry-level econometric analysis shows a moderate negative correlation ($r=-0.45$) between throughput and code quality (Section~\ref{sec:Risk:LimitCodeQuality}). These findings highlight the importance of organizational readiness and governance mechanisms that balance speed with quality through clear accountability and quality-assurance processes. Moreover, as shown in \ref{subsub:teamcollaboration}, reliance on LLM-assistants can alter communication patterns and reduce collaboration among developers. To sustain long-term benefits, organizations should embed LLM-assistants into workflows through well-defined policies and structured review procedures that ensure AI-generated artifacts receive rigorous oversight. As LLMs continue to advance in capability and autonomy, organizations should periodically revisit their adoption strategies and governance frameworks to ensure that new generations of tools strengthen, rather than disrupt, team collaboration, trust, and code quality.

\begin{mygraybox}
\textit{\textbf{Recommendation 4.}}  Organizations should assess their readiness for LLM-assisted development and adopt strategies that align productivity gains with code quality. This includes investing in training to cultivate calibrated trust, implementing quality-assurance processes for AI-generated artifacts, and fostering a culture that values collaboration and continuous learning alongside automation. \DIFadd{Given the documented negative correlation between throughput and quality~\cite{PS46}, organizations are advised to establish clear policies delineating which tasks are appropriate for LLM assistance, mandate enhanced review processes for LLM-generated code in high-risk modules, and monitor team collaboration patterns to detect and mitigate any erosion of knowledge sharing.}
\end{mygraybox}

\textbf{Professional and Ethical Considerations}. \DIFadd{The integration of LLMs into software engineering introduces complex professional and ethical challenges related to accountability, transparency, and fairness.} The black-box nature of proprietary LLM systems presents a transparency deficit. These tools frequently do not provide sources or references for their generated code, which hinders developers’ ability to conduct rigorous validation and understand the code’s lineage or security implications \cite{PS8, PS25}. This lack of transparency contributes to accountability gaps when LLM-generated code introduces defects, vulnerabilities, or potentially copyrighted material.



\begin{mygraybox} \DIFadd{
\textit{\textbf{Recommendation 5.}}  
Organizations should mandate explicit disclosure of AI-assisted contributions and establish clear accountability frameworks that define ownership, review requirements, and traceability of AI-generated artifacts. Developers must remain vigilant about algorithmic bias, as LLMs trained on skewed internet data can perpetuate unfair or discriminatory patterns. Ethical review and bias testing should therefore be integrated into standard software quality assurance processes, treating AI output as both a technical and socio-technical artifact. Finally, where feasible, practitioners should favor tools and workflows that improve transparency and traceability, enabling informed validation and responsible professional judgment.}
\end{mygraybox}

\subsection{Implications and recommendations for researchers }

\DIFadd{\textbf{Open issues and research gaps.}
Our synthesis identifies several gaps that require attention from the research community. First, developer well-being is not directly examined by any primary study, despite growing recognition that mental health and sustainable work practices are integral to long-term productivity. Second, human-human collaboration in LLM-mediated workflows remains understudied: only 3 of 10 studies addressing the Communication dimension, leaving unresolved questions regarding how LLM-assistants reshape pair programming, knowledge sharing, and collective code ownership. Third, long-term effects remain largely unknown, as the majority of studies employ short-term laboratory designs that cannot capture cumulative technical debt, sustained versus diminishing productivity gains, or skill erosion over extended periods. Fourth, the field lacks standardized metrics and validated instruments, which impedes cross-study comparison and cumulative knowledge building. Fifth, the conditions under which LLM-assistants improve versus degrade code quality remain unresolved, as this outcome is reported as both a benefit and a risk depending on study context and evaluation criteria.}

\textbf{Establishing shared practices and cumulative insight.}  
\DIFadd{The current body of research mainly rely on laboratory experiments, which represent the most common research strategy (38\%)}, focusing on controlled tasks to isolate effects. This is critical for early insights. However, the methodological diversity can challenge cross-study comparison and synthesis. For instance, LLM-assistant's effect on code quality is inconsistent, reported as both a benefit and a risk depending on study context and metrics used. Similarly, cognitive load findings are mixed: some studies report reduced mental effort, while others find no effect or increased frustration. These inconsistencies highlight the limits of the current evidence base and motivate context-aware, longitudinal studies that follow developers, teams, and organizations over time. Short-term studies alone cannot determine whether initial gains are sustained or whether they mask longer-term risks such as skill erosion, automation complacency, cumulative technical debt in evolving systems, or whether observed throughput translates into lasting organizational benefits.

\begin{mygraybox}
\textit{\textbf{Recommendation 1.}}
Researchers should adopt shared evaluation frameworks and validated instruments that allow comparability. Future work should include longitudinal, field, and team-based studies that capture long-term outcomes. \DIFadd{Specifically, future investigations should: (i) employ standardized cognitive load measures to reconcile the current inconsistency in findings; (ii) track productivity, code quality, and skill development over extended periods rather than single sessions; (iii) conduct organizational case studies that capture real-world complexity; and (iv) triangulate quantitative metrics with qualitative insights to elucidate the mechanisms underlying observed effects.}
\end{mygraybox}

\textbf{On the dimensions of productivity.} 
Existing work agrees that developer productivity is a multi-dimensional and context-sensitive construct \cite{petersen2011measuring}. This complexity is magnified in the context of LLM-assisted development. Our findings provide evidence that the vast majority of studies (\DIFadd{90}\%) examine at least two \textit{SPACE} dimensions. This indicates a positive move from the research community toward a multidimensional perspective. However, our findings also highlight that only \DIFadd{15}\% of studies extend beyond three dimensions, indicating that room for improvement remain in terms of providing a more complete evaluation of the productivity concept.

Satisfaction, Performance, and Efficiency are the most frequently investigated dimensions. Communication, however, remains underexplored with the human-human collaboration being the least investigated communication sub-dimension. This is inline with our findings on LLM-assistants risks on collaboration and further highlights the need to investigate the impact of LLM-assistants for both human-human and human-agent collaboration and
communication. As LLMs continue to advance, particularly in reasoning, and multimodal understanding, these dynamics may further evolve. Future studies should therefore examine how improvements in model capability reshape the balance between productivity, collaboration, and quality.

\begin{mygraybox}
\textit{\textbf{Recommendation 2.}}
Researchers should continue advancing multidimensional evaluations of developer productivity by systematically addressing underexplored \textit{SPACE} dimensions, particularly Communication and Collaboration. While Satisfaction, Performance, and Efficiency are frequently assessed, the human-human and human-agent interaction aspects remain limited. Future studies should incorporate richer measures of team dynamics and communication patterns to better understand how LLM-assistants affect collaborative workflows. \DIFadd{Priority areas for investigation include: (i) the effects of LLM adoption on pair programming and code review practices; (ii) developer well-being, occupational stress, and sustainable work practices; (iii) shifts in the allocation of developer time across tasks following LLM adoption; and (iv) the development and validation of instruments specifically designed for LLM-assisted development contexts.}
\end{mygraybox}

\textbf{Accounting for confounding variables.}
The variations in reported productivity outcomes across studies can often be attributed to confounding variables such as developer experience, task complexity, and domain context. These factors can impact the observed benefits and risks of LLM-assistants. For instance, novices may show higher immediate gains but also greater over-reliance and long-term skill erosion \cite{PS4, PS22, PS47}, while expert developers often achieve sustained efficiency through selective and critical use \cite{PS2, PS8, PS43}. Likewise, LLMs tend to perform well on isolated, well-defined tasks but struggle in complex, context-rich projects where validation overhead offsets speed gains \cite{PS8, PS25}. Current laboratory-based studies can have limited ecological validity. Future research should explicitly model and report these contextual variables to enhance fairer comparisons and more cumulative insight.

\begin{mygraybox}
\textit{\textbf{Recommendation 3.}}
Researchers should systematically account for confounding variables that influence productivity outcomes, including developer expertise, task complexity, and domain context. Study designs should incorporate these as covariates, ensuring that effects attributed to LLM-assistants reflect genuine productivity changes rather than contextual bias. Explicitly reporting these variables will strengthen cross-study comparability and help build a context-aware evidence base. \DIFadd{Essential methodological practices include: (i) stratifying analyses by experience level and reporting differential effects for novice versus expert developers; (ii) operationalizing task complexity along defined dimensions and examining its moderating role on LLM effectiveness; (iii) documenting organizational context to facilitate meta-analytic synthesis; and (iv) conducting replication studies across diverse populations, tools, and contexts to establish boundary conditions for reported findings.}
\end{mygraybox}



\section{Threats to Validity}
\label{sec:threats}

This systematic review acknowledges several threats to the validity of its findings, which arise both from the methodology employed and the evolving nature of the primary evidence base.
\subsection{Study Selection and Classification Rigor}
\textbf{Study selection bias.} A key threat lies in the potential omission of relevant studies due to the inclusion and exclusion criteria defined in our protocol (see Section~\ref{sec:inclusionexclusion}). Specifically,  we exclude shorter papers, non-peer-reviewed work, and publications outside academic journals and conference proceedings. To mitigate this, all authors collaboratively agree on these criteria to ensure methodological rigor and quality control to enhance the reliability of the synthesized findings. Another threat relates to the challenge of identifying the human-centered studies within the large body of LLM research. Our initial search strings include terms such as ``performance'' or ``efficiency'', which yielded results focused on the technical applications of LLMs rather than their impact on developer productivity. This issue stems from the broad scope of LLM4SE research \cite{hou2024large}, where such keywords frequently describe model behavior or algorithmic improvements rather than developer-centric outcomes. To mitigate this threat, three authors jointly review the selected control papers to ensure alignment with the inclusion criteria. We iteratively validate our search strings using control articles inspired by \textcite{zhang2011identifying} (see section~\ref{sec:query_formation}), ensuring all control papers are correctly retrieved. Furthermore, we employ backward and forward snowballing to further enhance the coverage of relevant articles and reduce the risk of missing key studies.

\noindent
\textbf{Bias and repeatability.}
The open-ended nature of our research questions can introduce the threat of taking subjective decisions in the interpretation and selection of studies. To mitigate this, we adopt a multi-step validation process involving all co-authors. While the initial screening and data extraction were conducted by one author, the remaining authors were actively involved in selection validation and protocol design. The teams held regular weekly meetings for a period of 9 months to discuss the inclusion decisions and refine the selection and data extraction process. Additionally, we employ a conservative screening strategy during the full-text review, whereby any study with uncertain eligibility was retained for further assessment and independently reviewed by two senior co-authors to minimize the risk of excluding relevant studies.

\noindent
\textbf{Classification rigor.}
A potential threat lies in the mapping of study findings to the \textit{SPACE} framework (see Section~\ref{sec:RQ3}). Since the \textit{SPACE} framework was not originally developed to represent productivity in human-LLM collaboration settings, assigning findings to specific sub-dimensions required interpretive decisions. This may have introduced subjective bias during data coding. To mitigate this, we adapted definitions from prior literature \cite{forsgren2021space, sikand2024much} to better fit the context of our review, and discussed coding decisions collaboratively in team meetings. 

\subsection{Limitations of the Primary Evidence Base}

\textbf{Formative and controlled studies}. A notable limitation inherited from the primary literature is the high proportion of formative \DIFadd{(59\%)} and laboratory experiment \DIFadd{(38\%)} studies (see section \ref{sec:RQ1}). While this emphasis provides strong internal validity (i.e., controlled effects), it can hinder the ecological validity of the findings. A number of reported productivity results largely reflect performance on isolated tasks and controlled environments. While these findings can overlook the complex realities of large-scale industry development, they are critical to building foundational knowledge.
\newline
\noindent
\textbf{Methodological diversity.} Methodological diversity across the primary studies can hinder comparability. The literature exhibits a lack of standardized metrics, particularly for highly contested concepts like code quality and cognitive load. For example, studies employ diverse measures, ranging from cyclomatic complexity and code coverage to defect density and functional correctness, to evaluate code quality. While this diversity can hinder comparability, it can also supports the triangulation of findings from different sources of evidence. 
 

\noindent
\textbf{Temporal relevance.} Given the rapid pace of progress in Generative AI, a key threat concerns the temporal completeness of the review. Our search and data extraction were conducted at the end of 2024, with \DIFadd{77\%} of the included studies published in the same year. As LLM capabilities, evaluation methods, and developer practices evolve, new findings may emerge quickly after our review period. To mitigate this limitation, we emphasize the importance of transparent, replicable protocols and encourage periodic updates that can incorporate emerging empirical evidence in this fast-evolving field.

\section{Conclusion }
\label{sec:conclusion}

In this paper, we investigate LLM-assistants' impact on developer productivity. To achieve this, we systematically identify and analyze \DIFadd{39} peer-reviewed studies from their methodological strategies, evaluation practices, and the productivity dimensions these studies focus on. We synthesize reported benefits and risks, and apply established conceptual frameworks to map our findings and contextualize their broader implications. 

Our analysis reveals a range of reported benefits, including reduced task initiation overhead, accelerated development, and support for code-adjacent tasks. At the same time, studies identify several risks, such as over-reliance on LLM-assistants especially affecting novice programmers, disruptions to developer flow, and reduced team communication or collaboration. Code quality, in particular, has a mixed outcome, with studies reporting both improvements and degradations depending on context, task design, and evaluation criteria. Our findings show that most studies \DIFadd{(90\%)} consider multiple productivity dimensions. However, relatively few studies extend beyond three, and dimensions like communication and, more specifically, human-human collaboration remain underexplored. 

Looking ahead, there is value in expanding the evidence base through team-based studies that capture the dynamic and socio-technical nature of software development. As LLM-assistants become more deeply embedded in everyday workflows, future research will play a critical role in understanding the multidimensional impact of LLM-assistants on developer productivity. To facilitate transparency and future work, we provide a publicly available replication package \cite{anonymous2025replication}.



\section{Acknowledgment}
This work was supported by NSERC Discovery Grant RGPIN-2024-06511.


\printbibliography


\end{document}